\documentclass[a4paper,nofootinbib]{revtex4}
\usepackage{graphicx}

\begin{document}

\title{Thermoelectric Power and Hall Effect in Correlated Metals and Doped
Mott -- Hubbard Insulators: DMFT approximation}

\author{E.Z. Kuchinskii, N.A. Kuleeva, M.V.Sadovskii}

\affiliation{Institute for Electrophysics, Russian Academy of Sciences,
Ural Branch, Amundsen str. 106, Ekaterinburg 620016, Russia}


\begin{abstract}
We present comparative theoretical investigation of thermoelectric
power and Hall effect in the Hubbard model for correlated metal and Mott
insulator (considered as prototype cuprate superconductor) for different
concentrations of current carriers. Analysis is performed within standard
DMFT approximation. For Mott insulator we consider the typical case of partial
filling of the lower Hubbard band (hole doping).
We calculate the dependence of thermopower on doping level and determine the
critical concentration of carriers corresponding to sign change of thermopower.
An anomalous dependence of thermopower on temperature is obtained significantly
different from linear temperature dependence typical for the usual metals.
The role of disorder scattering is analyzed on qualitative level.
The comparison with similar studies of the Hall effect shows, that breaking of
electron -- hole symmetry leads to the appearance of the relatively large
interval of band -- fillings (close to the half -- filling) where thermopower
and Hall effects have different signs. We propose a certain scheme allowing to
determine the number of carriers from ARPES data and perform semi --
quantitative estimate of both thermopower and Hall coefficient using the usual
DFT calculations of electronic spectrum.
\end{abstract}

\pacs{71.10.Fd, 74.20.-z, 74.20.Mn}

\maketitle


\section{Introduction}

The problem of doping dependence of thermopower and Hall effect in strongly
correlated systems is known for a long time. It remains actual with respect to
experimental studies of these effects in copper -- oxide (cuprate)
high -- temperature superconductors at different doping levels.
One of the basic concepts in cuprate physics is that these systems are strongly
correlated and their metallic (and superconducting) state is realized via
doping of the prototype phase of the Mott insulator, which in the simplest
case can be described in the framework of the Hubbard model.

Most developed method of theoretical description of the Hubbard model for the
wide range of parameters of this model remains the dynamical mean -- field
approach (DMFT) \cite{pruschke,georges96,Vollh10}. Systematic studies of
concentration and temperature dependencies of the Hall effect was performed in
our recent papers \cite{KKKS1,KKKS2}. Similar papers on concentration and
temperature dependencies of the empower are practically absent. This paper is
devoted to the study of this problem in comparison with previous results on
the Hall effect.

\section{Thermopower and Hall coefficient. General relations.}

As noted above the general method to study the Hubbard model is the dynamical
mean -- field theory (DMFT) \cite{pruschke,georges96,Vollh10}, which gives an
exact description of the system in the limit of infinite dimensions (lattice
with an infinite number of nearest neighbors). Going beyond this approximation
\cite{GDMFT, RMP} is usually much more complicated. The aim of the present work
is the systematic study of concentration and temperature dependence of
thermopower at different dopings of the lower Hubbard band within the standard
DMFT approximation. We also compare in detail the results for thermopower with
similar results obtained in our previous papers using the similar approach
\cite{KKKS1,KKKS2}.

Anticipating the possible comparison with experimental data for cuprates in the
following we consider the two -- dimensional model of electronic spectrum in
tight -- binding approximation:
\begin{equation}
\varepsilon ({\bf p})=-2t(cos(p_xa)+cos(p_ya))-4t'cos(p_xa)cos(p_ya),
\label{SPtt'}
\end{equation}
where $a$ -- is (square) lattice parameter.

For this two -- dimensional model in the following we shall consider a number
of typical cases:
\begin{enumerate}
\item{spectrum with hopping only between nearest neighbors ($t'=0$) and complete
electron -- hole symmetry,}
\item{spectrum with $t'/t=-0.25$ qualitatively corresponding to electron
in systems like LSCO,}
\item{spectrum with $t'/t=-0.4$, qualitatively corresponding to situation
observed in YBCO.}
\end{enumerate}

Thermopower is determined by the following expression
\cite{pruschke, Madelung, Phillips2010}:
\begin{equation}
S=-\frac{k_B}{e}\frac{1}{T}\frac
{\int_{-\infty}^{\infty}d\varepsilon \varepsilon \tau (\varepsilon)
\left( -\frac{\partial f(\varepsilon)}{\partial \varepsilon} \right)}
{\int_{-\infty}^{\infty}d\varepsilon \tau (\varepsilon)
\left( -\frac{\partial f(\varepsilon)}{\partial \varepsilon} \right)}.
\label{S1}
\end{equation}
where $k_{B}$ is Boltzmann's constant (in the following we write temperature
in the units of energy assuming $ k_{B}=1$), $e$ is electronic charge,
$f(\varepsilon)={(e^{\frac{\varepsilon}{T}}+1})^{-1}$ -- Fermi distribution.
Below the values of thermopower are shown in units of
$\frac{k_B}{e} \approx 86 \left [\frac{\mu V}{K}\right]$.

Relaxation parameter introduced in Eq. (\ref{S1}) is given by:
\begin{equation}
\tau(\varepsilon)=\sum_{{\bf p}\sigma}\left( \frac{\partial
\varepsilon ({\bf p})}{\partial p_x} \right) ^2
A^2({\bf p}\varepsilon)
\label{tau1}
\end{equation}
and is completely determined by Green's function $G({\bf p}\varepsilon)$
spectral density:
\begin{equation}
A({\bf p}\varepsilon)=-\frac{1}{\pi}ImG^R({\bf p}\varepsilon).
\label{SpDens}
\end{equation}
In standard \cite{pruschke,georges96,Vollh10} self -- energy of Green's function
is local, i.e. independent of momentum. Due to this locality both the usual and
Hall conductivities are also completely determined by the spectral density
$A({\bf p}\varepsilon)$.

The usual (diagonal) static conductivity is given by \cite{pruschke}:
\begin{equation}
\sigma_{xx}=\frac{\pi e^2}{2\hbar}\int_{-\infty}^{\infty}d\varepsilon
\left( -\frac{\partial f}{\partial \varepsilon} \right)\tau(\varepsilon),
\label{Gxx1}
\end{equation}
while Hall (non -- diagonal) conductivity determined by:
\begin{equation}
\sigma^H_{xy}=-\frac{2\pi^2e^3a^2H}{3\hbar^2}\int_{-\infty}^{\infty}
d\varepsilon
\left( -\frac{\partial f}{\partial \varepsilon} \right)\tau_{H}(\varepsilon),
\label{Gxy1}
\end{equation}
where  $H$ is the magnetic field along $z$ -- axis. Here we also introduced
Hall relaxation parameter:
\begin{equation}
\tau_{H}=\sum_{{\bf p}\sigma}\left( \frac{\partial
\varepsilon ({\bf p})}{\partial p_x} \right) ^2 \frac{\partial^2
\varepsilon ({\bf p})}{\partial {p_y}^2}
A^3({\bf p}\varepsilon).
\label{tauH1}
\end{equation}
Thus, the Hall coefficient is:
\begin{equation}
R_H=\frac{\sigma^H_{xy}}{H\sigma_{xx}^2}
\label{R_H1}
\end{equation}
and it is also completely determined by spectral density $A({\bf p}\varepsilon)$,
which we shall calculate within DMFT. Effective single Anderson impurity
model of DMFT in this work, as in Refs. \cite{KKKS1,KKKS2}, will be solved
using the numerical renormalization group approach (NRG) (NRG) \cite{NRGrev}.

Consider the case of low temperatures ($T\ll E_F$).
For any function $\Phi(\varepsilon)$ (if integral is converges) we have:
\begin{equation}
\int_{-\infty}^{\infty}d\varepsilon \Phi(\varepsilon)
\left( -\frac{\partial f}{\partial \varepsilon} \right)=\Phi(0)
+\frac{\pi^2}{6}T^2
\left( \frac{\partial^2 \Phi}{\partial \varepsilon^2} \right)_{\varepsilon=0}+...
\label{F}
\end{equation}
Then for diagonal conductivity from Eq. (\ref{Gxx1}) we get:
\begin{equation}
\sigma_{xx}=\frac{e^2}{\hbar}\frac{\pi}{2}\tau(0),
\label{Gxx2}
\end{equation}
and thermopower at low temperatures is given by:
\begin{equation}
S=-\frac{k_B}{e}\frac{\pi^2}{3}T\left.{\frac{\frac{d\tau(\varepsilon)}
{d\varepsilon}}{\tau(\varepsilon)}}\right|_{\varepsilon=0}=
-\frac{k_B}{e}\frac{\pi^2}{3}T\left.{\frac{d\ln \tau(\varepsilon)}
{d\varepsilon}}\right|_{\varepsilon=0},
\label{S2}
\end{equation}
so that from Eqs. (\ref{S2}) and (\ref{Gxx2}) we immediately obtain Seebeck's
expression \cite{Madelung, Ziman}:
\begin{equation}
S=-\frac{k_B}{e}\frac{\pi^2}{3}T\frac{d\ln \sigma_{xx}(\mu)}{d\mu}.
\label{S3}
\end{equation}
where $\mu$ is the chemical potential (from which we always count the energy
$\varepsilon$).

Thus at low $T$ the absolute value of thermopower linearly increases with the
growth of $T$, while the sign of $S$ is completely determined by the sign of
$\left.{\frac{d\tau(\varepsilon)}{d\varepsilon}}\right|_{\varepsilon=0}$
or $\frac{d\sigma_{xx}(\mu)}{d\mu}$. In the following we shall be interested in
dependence on band -- filling
$n=n_{\uparrow}=n_{\downarrow}=\int_{-\infty}^{\infty}d\varepsilon
f(\varepsilon)N(\varepsilon)$
(we shall consider only the paramagnetic state), where $N(\varepsilon)$ is the
density of states per single spin projection. If we explicitly take into account
the chemical potential Eq. (\ref{S2}) can be rewritten as:
\begin{equation}
S=-\frac{k_B}{e}\frac{\pi^2}{3}T\frac{\frac{d\tau (\mu)}{d\mu}}{\tau (\mu)}=
-\frac{k_B}{e}\frac{\pi^2}{3}T\frac{dn}{d\mu}\frac{\frac{d\tau (n)}{dn}}
{\tau (n)}.
\label{S4}
\end{equation}
Here $\frac{dn}{d\mu}>0$ as at low temperatures $\frac{dn}{d\mu}=\frac{d}{d\mu}
\int_{-\infty}^{\mu}d\varepsilon N(\varepsilon)=N(\mu)>0$.
Thus, the sign of $S$ is completely determined by the sign of
$\frac{d\tau(n)}{dn}$ or $\frac{d\sigma_{xx}(n)}{dn}$.

Hall conductivity (\ref{Gxy1}) at low temperatures with the account of
Eq. (\ref{F}) takes the form:
\begin{equation}
\sigma^H_{xy}=-\frac{2\pi^2e^3a^2H}{3\hbar^2}\tau_{H}(0)
\label{Gxy2}
\end{equation}
and the Hall coefficient
\begin{equation}
R_H=-\frac{a^2}{e}\frac{8}{3}\frac{\tau_{H}(0)}{\tau^2(0)}.
\label{R_H2}
\end{equation}
The sign of the Hall coefficient is completely determined by the sign of
$\tau_{H}(0)$ or by the sign of $\tau_{H}(n)$.
We see that the signs of thermopower and Hall coefficient are determined,
strictly speaking, by completely different expressions, so that the
band -- fillings corresponding to their sign change, in general, can be quite
different.

\section{Thermopower and Hall coefficient in the absence of correlations ($U=0$).}

To study this situation in more details we shall consider first the case when
electron correlations are absent (Hubbard interaction $U=0$.)  In the absence
correlations and other scattering processes of electrons (ideal conductor)
both numerator and denominator in Eq. (\ref{S1}) for thermopower and in
Eq. (\ref{R_H1}) for the Hall coefficient diverge, thus to regularize our
calculations we have to introduce some weak electron scattering, taking
the single -- electron Green's as:
\begin{equation}
G^R({\bf p}\varepsilon)=\frac{1}{\varepsilon-\varepsilon({\bf p})+i\gamma},
\label{GR}
\end{equation}
where $\gamma \ll t$ is scattering frequency (e.g. by impurities), so that
the spectral density takes the form:
\begin{equation}
A({\bf p}\varepsilon)=\frac{1}{\pi}
\frac{\gamma}{(\varepsilon-\varepsilon({\bf p}))^2+\gamma^2}.
\label{SpDens_g}
\end{equation}

\begin{figure}
\includegraphics[clip=true,width=0.48\textwidth]{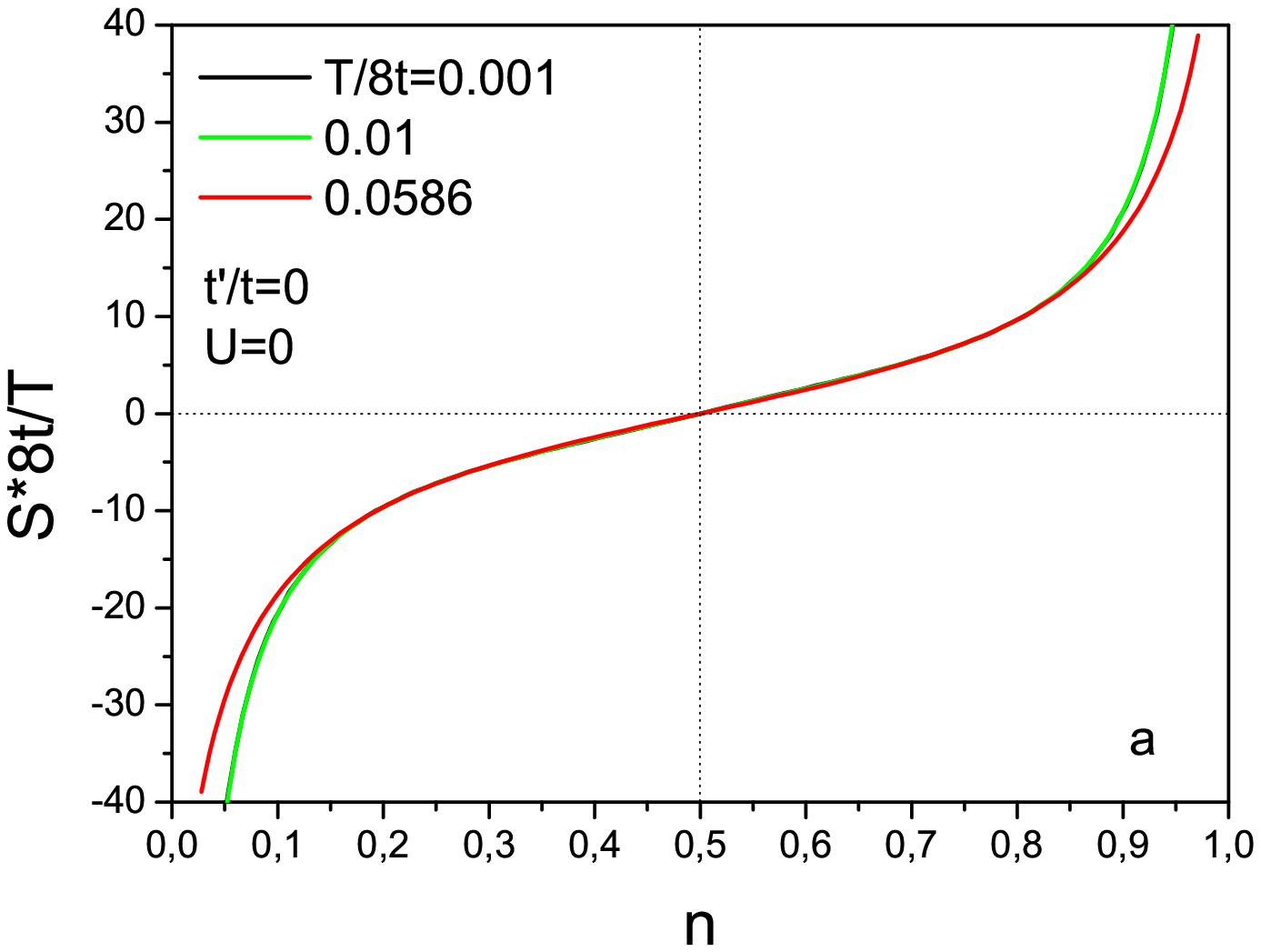}
\includegraphics[clip=true,width=0.48\textwidth]{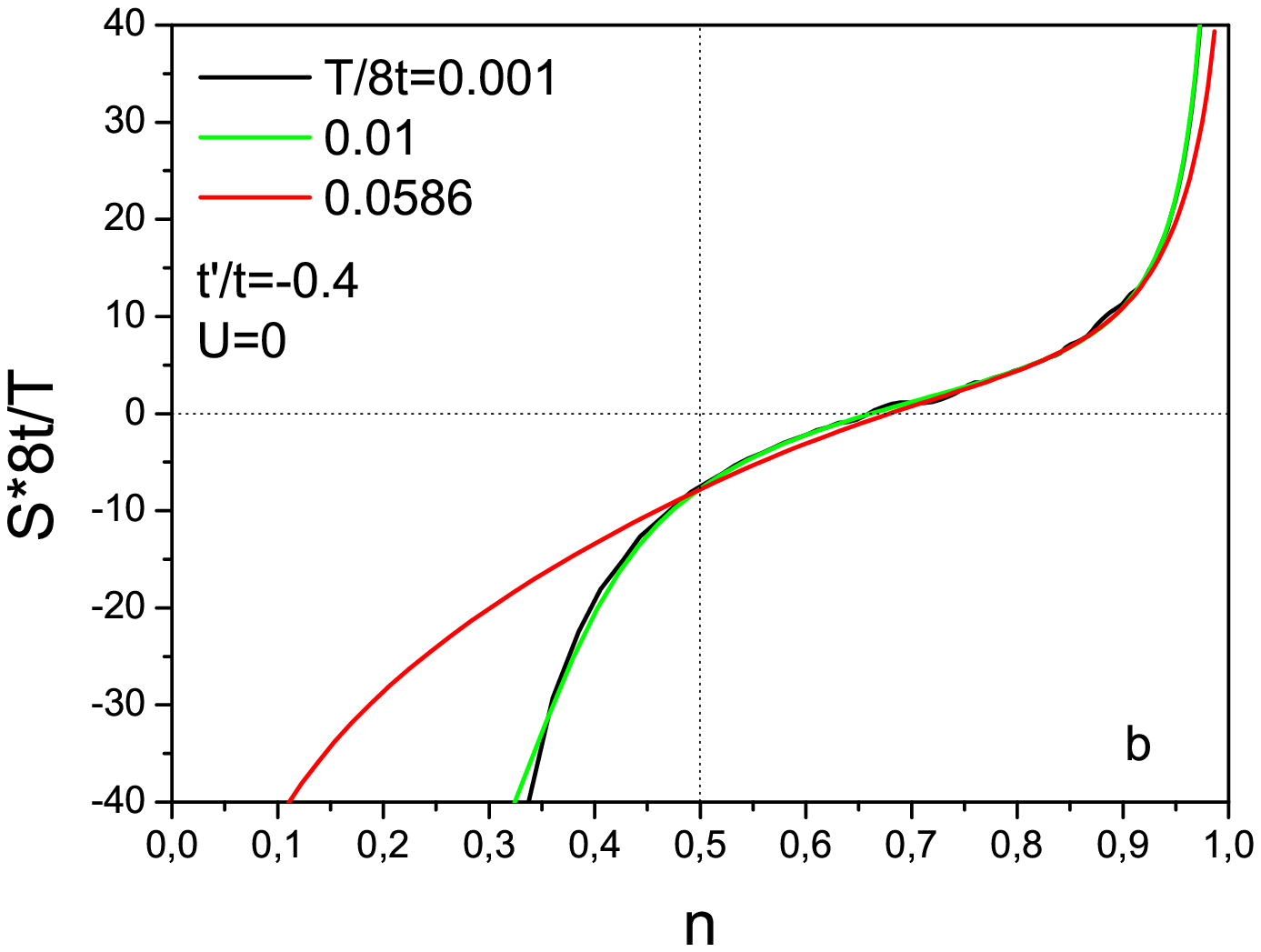}
\includegraphics[clip=true,width=0.48\textwidth]{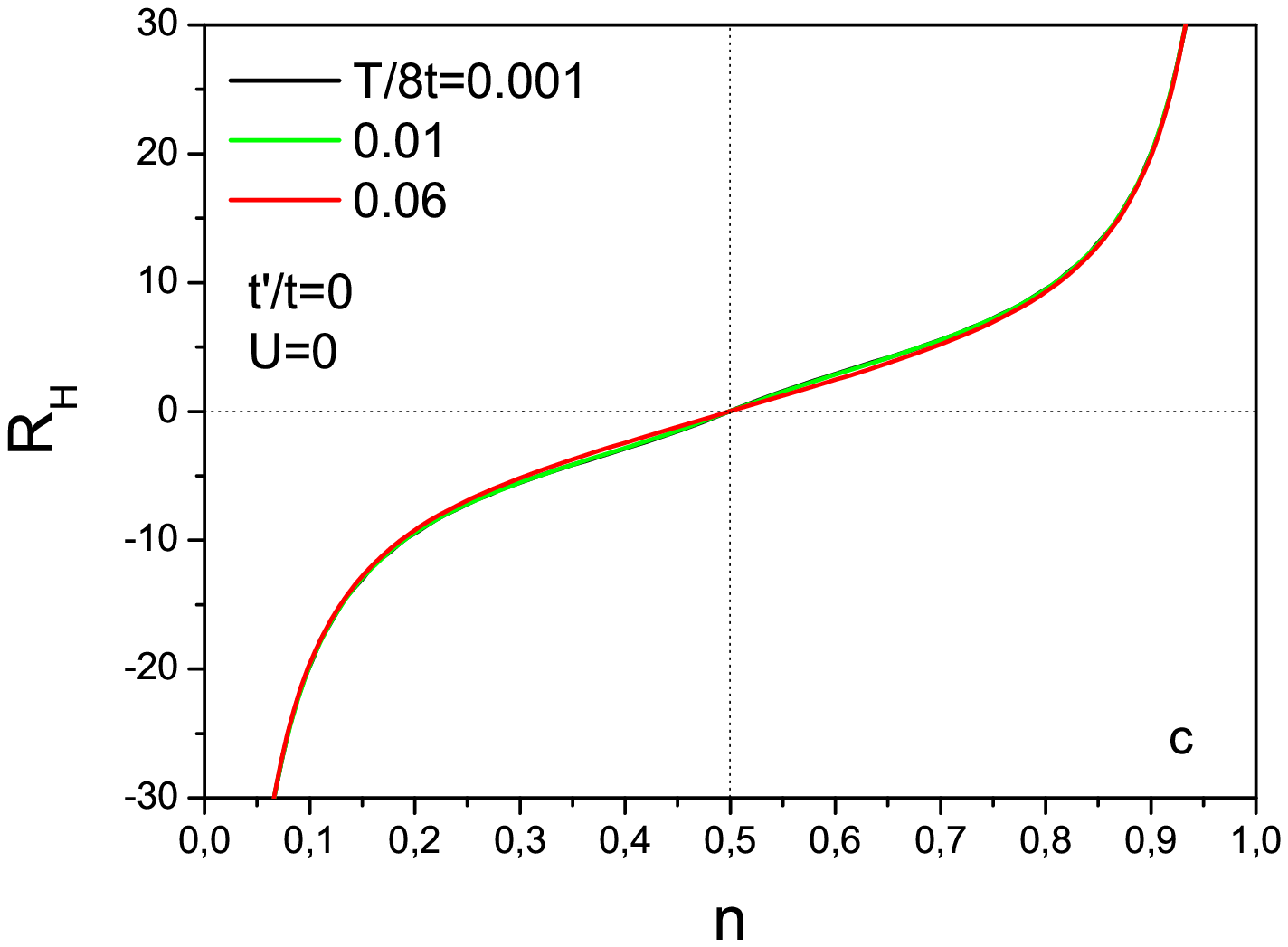}
\includegraphics[clip=true,width=0.48\textwidth]{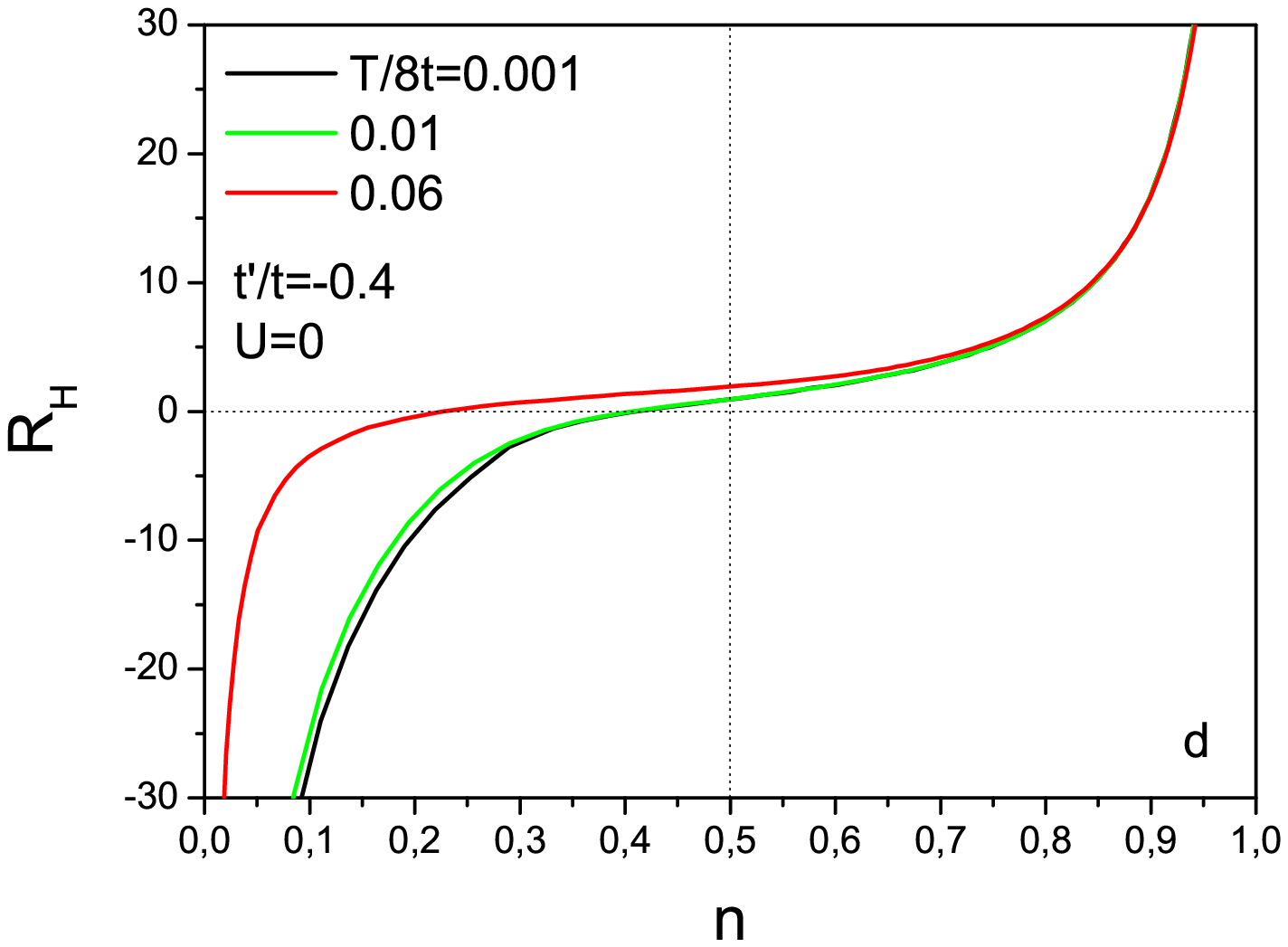}
\caption{Dependence of thermopower (a,b) and Hall coefficient (c,d) on
band -- filling for $t'=0$ -- left column (a,c)
and $t'/t=-0.4$ -- right column (b,d).}
\label{fig1}
\end{figure}

Dependencies of thermopower and Hall coefficient obtained directly from Eqs.
(\ref{S1}) and (\ref{R_H1}) using the spectral density from Eq. (\ref{SpDens_g})
with $\gamma/8t=0.005$, are shown in Fig. \ref{fig1}, both for the case of
complete electron -- hole symmetry ($t'=0$) and for $t'/t=-0.4$, typical e.g.
for cuprates like YBCO. We can see that in the case of complete electron --
hole symmetry thermopower (Fig. \ref{fig1}a) is linear in temperature in
accordance with Eqs. (\ref{S2}) and (\ref{S3}), up to high temperatures
$T/8t\approx 0.06$, while $R_H$ (Fig. \ref{fig1}c) is practically independent
of $T$. Both thermopower and Hall coefficient change sign at band half --
filling. In case of pretty strong breaking of such symmetry ($t'/t=-0.4$) we
observe a noticeable deviation from linear dependence of thermopower on
temperature (cf. Fig. \ref{fig1}b), at high $T/8t\approx 0.06$, and Hall
coefficient (Fig. \ref{fig1}d) at high $T$ also acquires a noticeable dependence
on temperature. The sign change of $S$ is observed at filling $n\approx 0.65$,
which is significantly larger than half -- filling, while the Hall coefficient
changes its sign at band -- filling noticeably lower than half -- filling,
and with the growth of temperature this deviation from half -- filling further
grows and for $T/8t\approx 0.06$ $R_H$ changes sign at $n\approx 0.22$.
Thus for $t'/t=-0.4$ we observe rather wide region of band -- fillings $n$,
where the Hall coefficient $R_H$ already has positive sign (hole -- like),
while the thermopower $S$ is still negative (electron -- like).
As was shown above the sign of thermopower is completely determined by the sigh
of $\frac{d\tau(n)}{dn}$, while the sigh of Hall coefficient is determined by
the sign of $\tau_{H}(n)$. Thus, to clarify situation it may be useful to
analyze the dependencies of $\tau(n)$ and $\tau_{H}(n)$. But both relaxation
parameter $\tau$ and Hall relaxation parameter $\tau_{H}$ diverge as
$\tau\sim\frac{1}{\gamma}$ and $\tau_{H}\sim\frac{1}{\gamma^2}$ for
$\gamma \rightarrow 0$, so that it is useful to introduce some reduced
relaxation parameters, which are independent of $\gamma$, but are
characterized by some characteristics of spectrum at the Fermi surface.
We shall see that such parameters will be useful also for the analysis of
systems with strong electronic correlations.

For $\gamma \ll t$ spectral density has a narrow peak at
$\varepsilon \sim \varepsilon({\bf p})$, then:
\begin{eqnarray}
\tau(\varepsilon)=\sum_{{\bf p}\sigma}\left(
 \frac{\partial \varepsilon ({\bf p})}{\partial p_x} \right)^2
\left( \frac{1}{\pi}\frac{\gamma}{(\varepsilon-\varepsilon ({\bf p}))^2
+{\gamma}^2} \right) ^2=\nonumber\\
=\int_{-\infty}^{\infty}d\xi \left[\sum_{{\bf p}\sigma}\left(
\frac{\partial\varepsilon ({\bf p})}{\partial p_x} \right)^2 \delta
(\xi-\varepsilon ({\bf p}))\right]\times\nonumber\\
\times\left( \frac{1}{\pi}\frac{\gamma}{(\varepsilon-\xi)^2+{\gamma}^2} \right)^2
\label{tau2}
\end{eqnarray}
so that we can write:
\begin{equation}
\tau(\varepsilon)\approx \tau_{0}(\varepsilon)
\int_{-\infty}^{\infty}d\xi \left(\frac{1}{\pi}\frac{\gamma}
{(\varepsilon-\xi)^2+\gamma^2}\right) ^2=\frac{1}{2\pi\gamma}
\tau_{0}(\varepsilon),
\label{tau3}
\end{equation}
where we have introduced
\begin{equation}
\tau_{0}(\varepsilon)=\sum_{{\bf p}\sigma}\left(
\frac{\partial\varepsilon({\bf p})}{\partial p_x}\right)^2\delta(
\varepsilon-\varepsilon({\bf p})).
\label{tau0}
\end{equation}
Similarly
\begin{eqnarray}
\tau_{H}(\varepsilon)=\sum_{{\bf p}\sigma}
\left(\frac{\partial\varepsilon({\bf p})}{\partial p_x}\right)^2
\frac{\partial^2\varepsilon({\bf p})}{{\partial p_y}^2}
A^3(\varepsilon,{\bf p})\approx\nonumber\\
\approx\tau_{0H}(\varepsilon)
\int_{-\infty}^{\infty}d\xi \left(\frac{1}{\pi}
\frac{\gamma}{\xi^2+\gamma^2}\right)^3=
\frac{3}{8}\frac{1}{\pi^2\gamma^2}\tau_{0H}(\varepsilon),
\label{tauH2}
\end{eqnarray}
where we have introduced
\begin{equation}
\tau_{0H}(\varepsilon)=\sum_{{\bf p}\sigma}\left(
\frac{\partial\varepsilon({\bf p})}{\partial p_x}\right)^2
\frac{\partial^2\varepsilon({\bf p})}{\partial{p_y}^2}
\delta(\varepsilon-\varepsilon({\bf p})).
\label{tau0H}
\end{equation}
Then at low temperature ($T\ll E_F$) instead of Eq. (\ref{S4}) we obtain for the
thermopower:
\begin{equation}
S=-\frac{k_B}{e}\frac{\pi^2}{3}T\frac{\frac{d\tau_{0}(\mu)}{d\mu}}
{\tau_{0}(\mu)}=
-\frac{k_B}{e}\frac{\pi^2}{3}T\frac{dn}{d\mu}\frac{\frac{d\tau_{0}(n)}{dn}}
{\tau_{0}(n)}
\label{S5}
\end{equation}
and the sign of $S$ is opposite to the sign of $\frac{d\tau_{0}(n)}{dn}$.

For Hall coefficient instead of Eq. (\ref{R_H2}) we obtain
\begin{equation}
R_H=-\frac{a^2}{e}\frac{4\tau_{0H}(n)}{\tau_0^2(n)}
\label{R_H3}
\end{equation}
and the sign of $R_H$ is opposite to the sign of $\tau_{0H}(n)$. Note that
$\frac{\partial^2\varepsilon({\bf p})}{\partial{p_y}^2}$
is the only sign -- changing entity in Eq. (\ref{tau0H}).

\begin{figure}
\includegraphics[clip=true,width=0.48\textwidth]{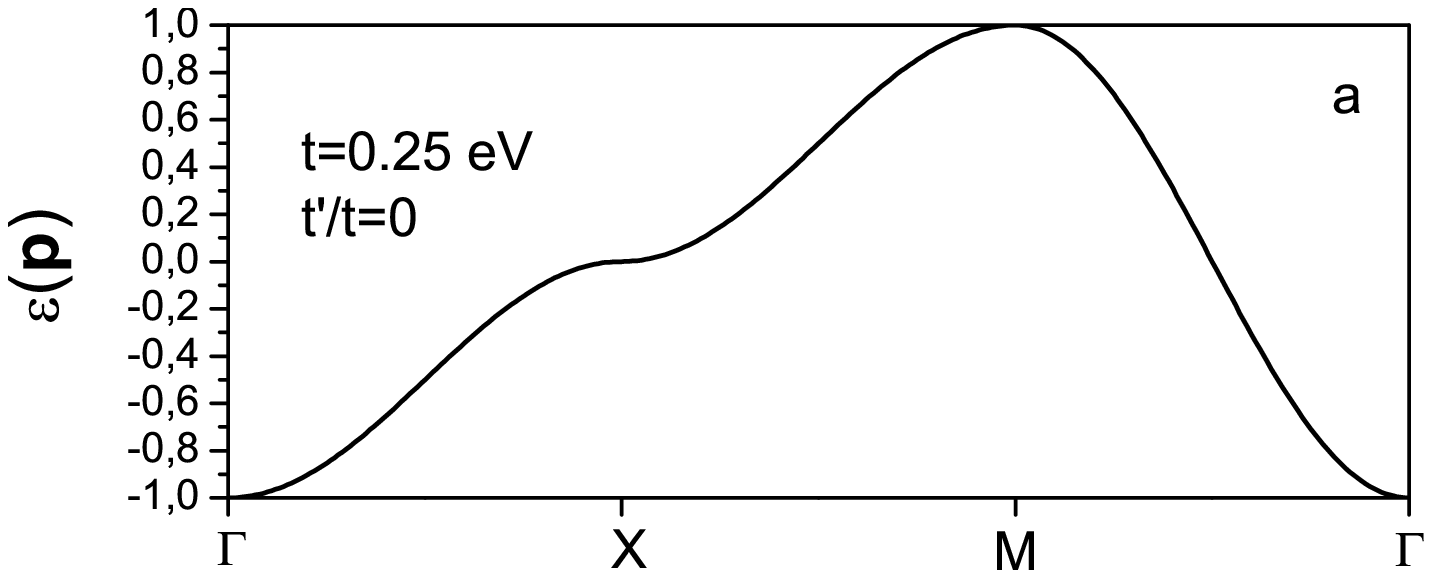}
\includegraphics[clip=true,width=0.48\textwidth]{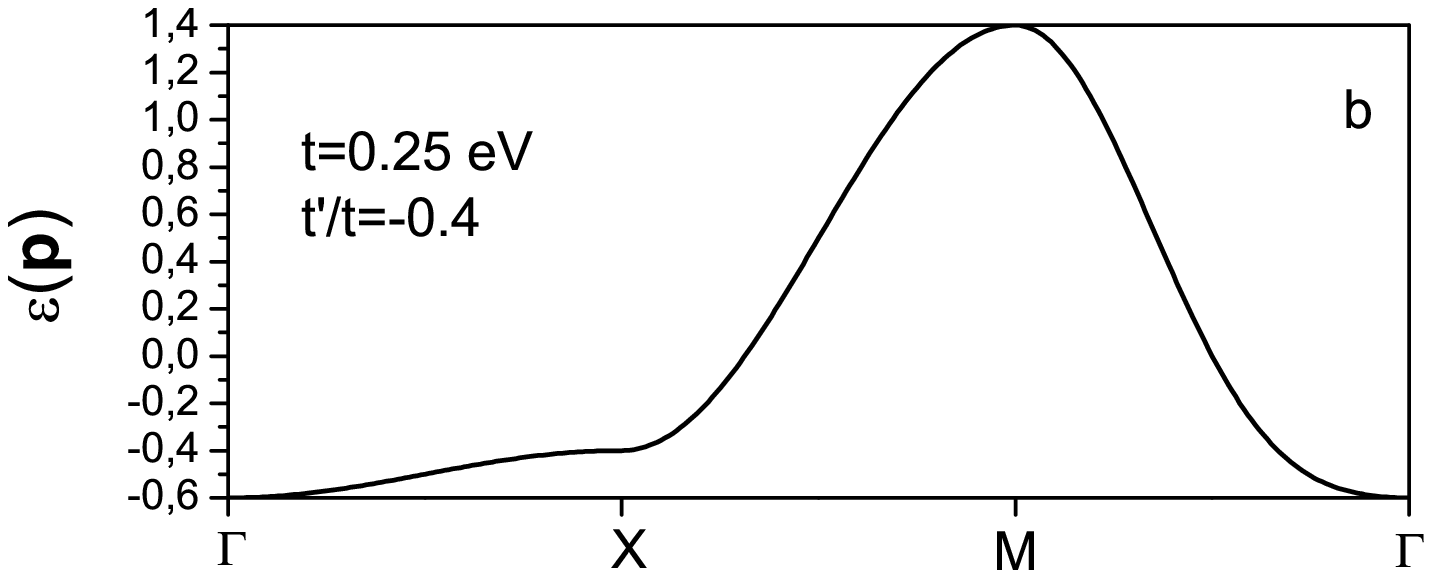}
\includegraphics[clip=true,width=0.48\textwidth]{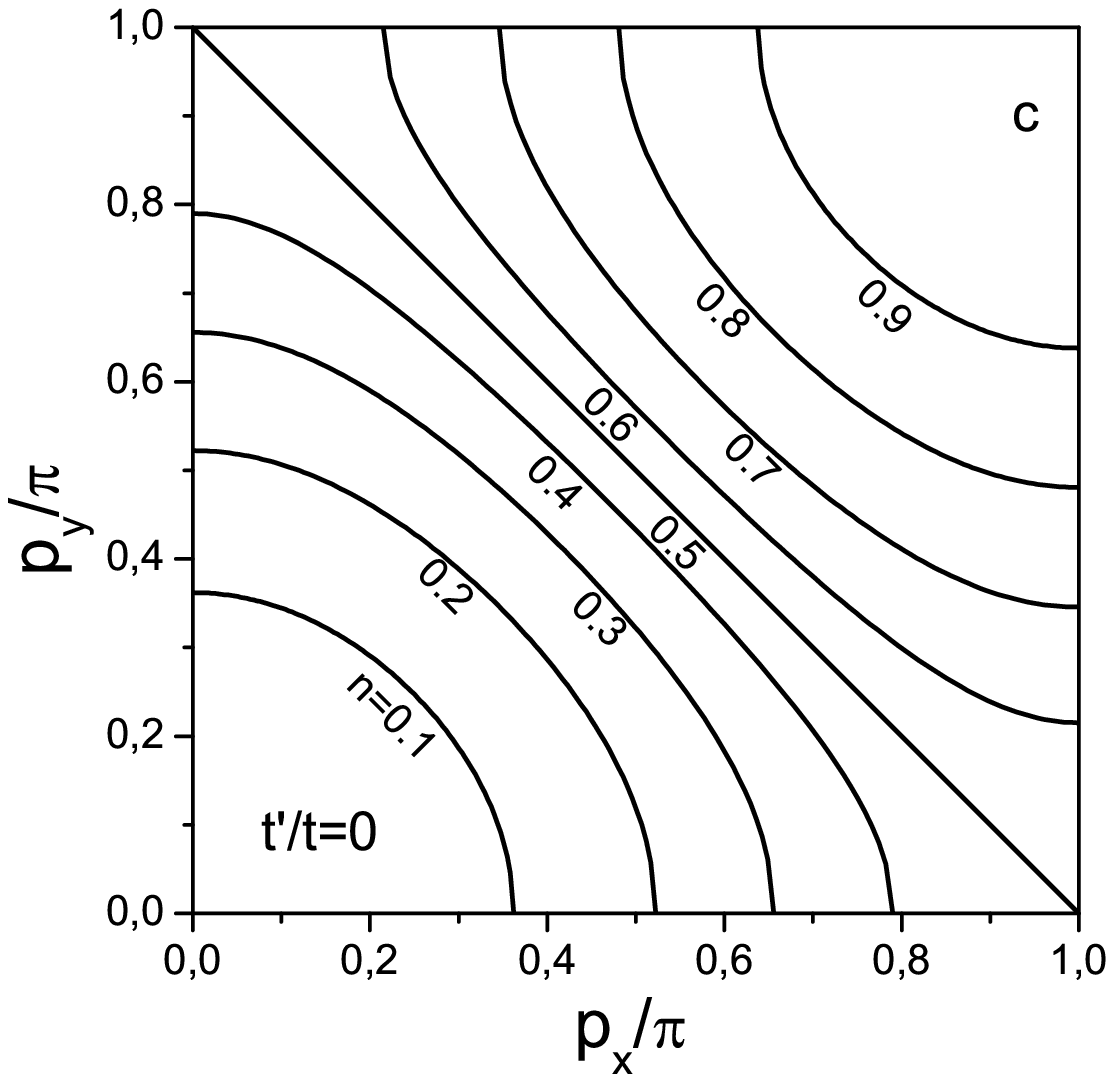}
\includegraphics[clip=true,width=0.48\textwidth]{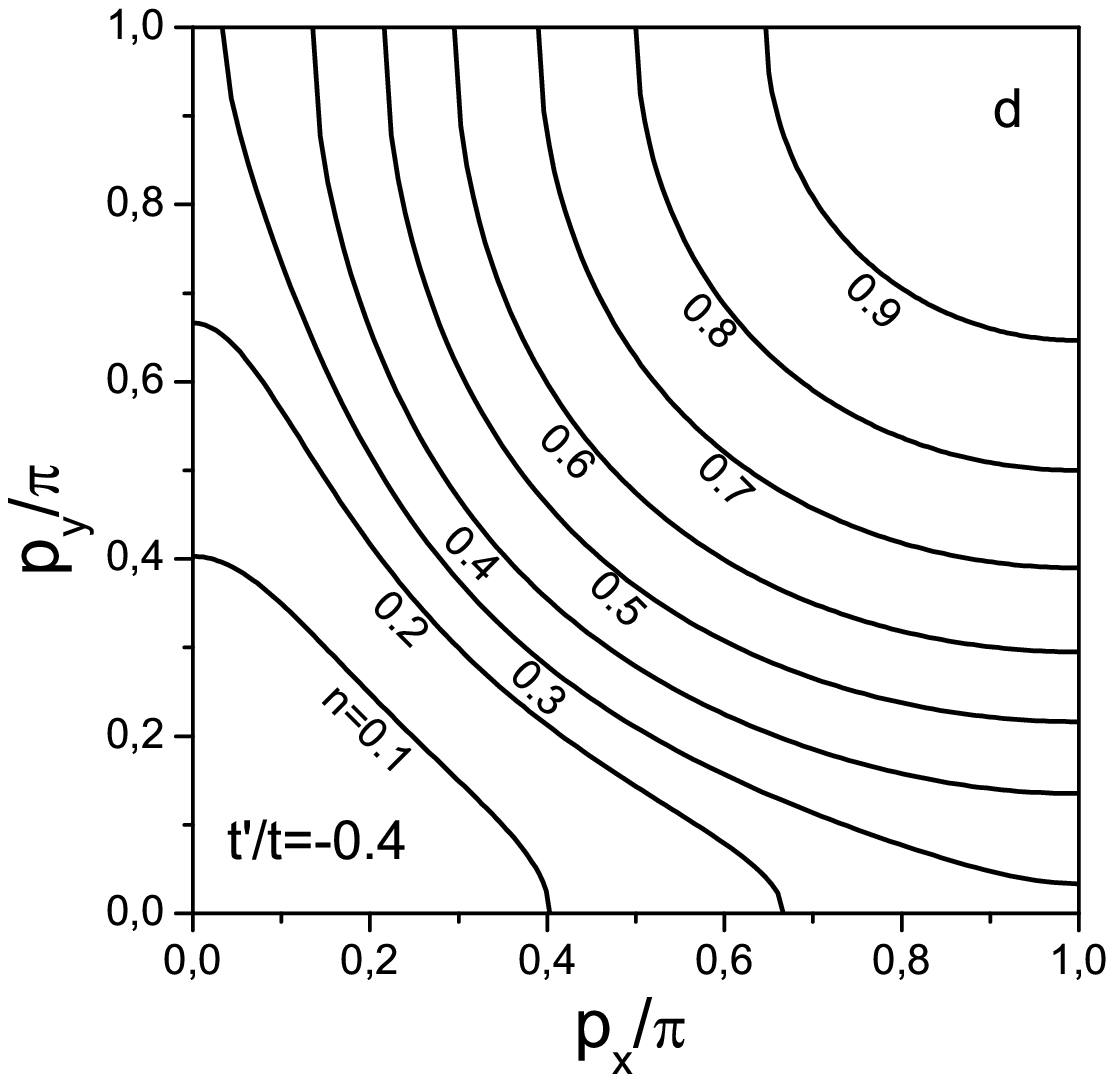}
\includegraphics[clip=true,width=0.48\textwidth]{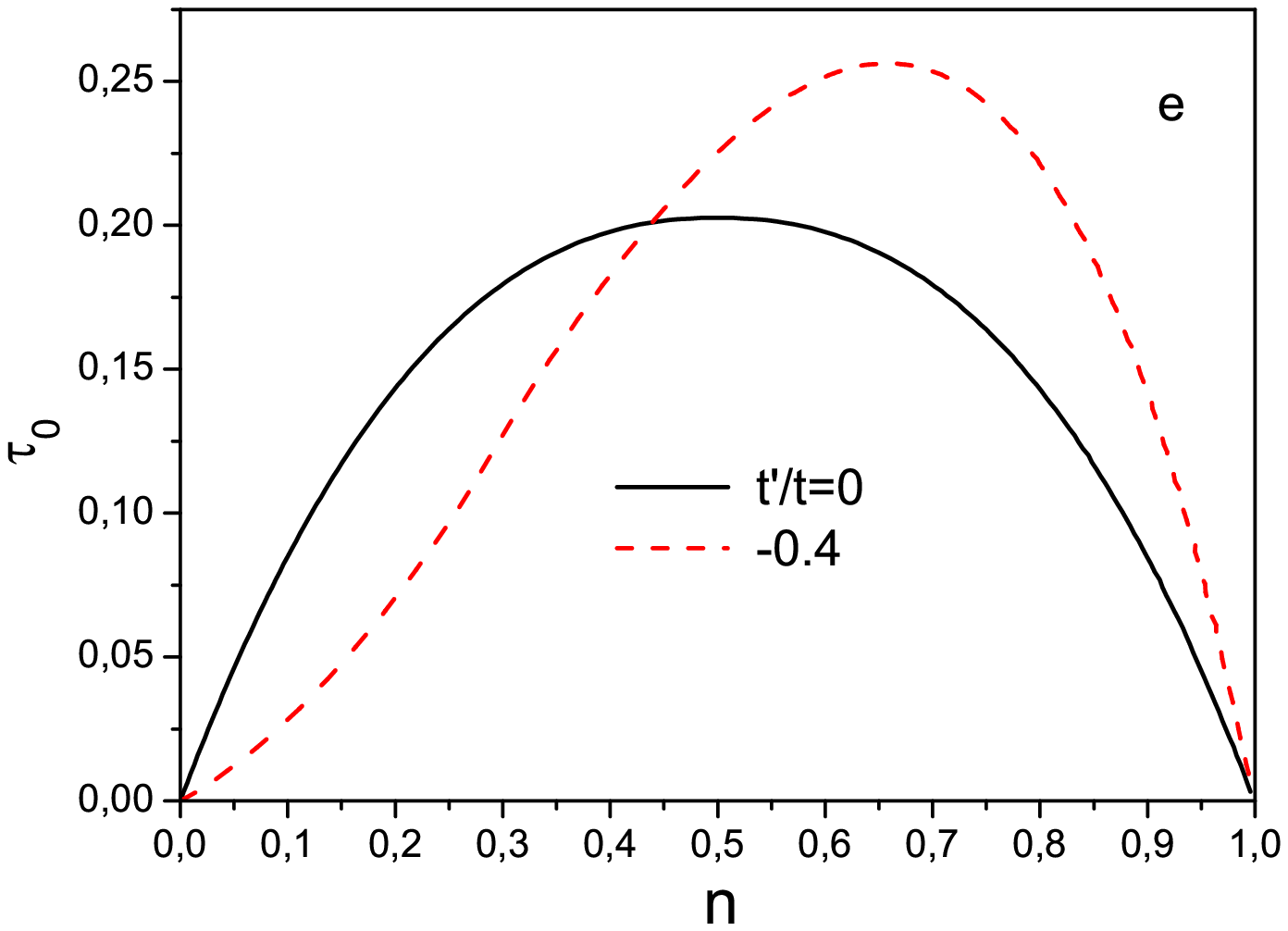}
\includegraphics[clip=true,width=0.48\textwidth]{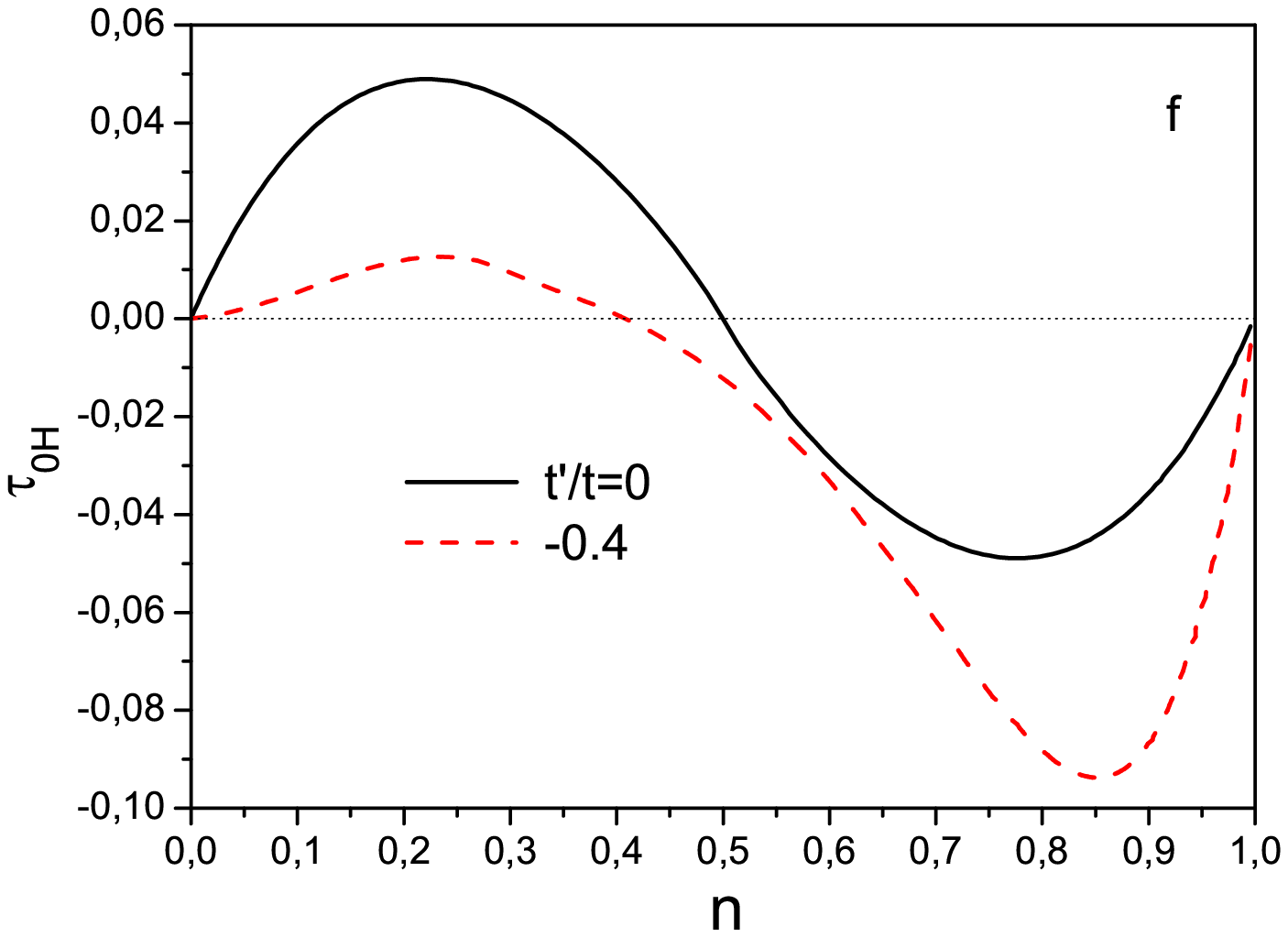}
\caption{Electron spectrum along symmetry directions in the Brillouin zone (a,b),
Fermi surfaces for different band -- fillings for (c,d) and dependences of
reduced relaxation parameters $\tau_0$ and $\tau_{0H}$ on band -- filling (e,f)
for $t'=0$,  $t'/t=-0.4$.}
\label{fig2}
\end{figure}

In Fig. \ref{fig2} we show dispersions $\varepsilon({\bf p})$ along symmetry
directions in Brillouin zone (Fig. \ref{fig2}a,b) and Fermi surfaces,
corresponding to different band -- fillings (Fig. \ref{fig2}c,d) for two
choices of $t'/t$. In the case of complete electron -- hole symmetry ($t'=0$)
for half -- filled band we observe the change of the type of Fermi surface
(Fig. \ref{fig2}c) from electron pocket around $\Gamma$ -- point of the
Brillouin zone, characteristic for $n<0.5$, to hole pocket around M -- point of
the Brillouin zone, characteristic for $n>0.5$. Due to electron -- hole symmetry
the regions of quadratic electron spectrum (close to $\Gamma$ --
point) and similar hole spectrum (close to M - point) are just the same
(Fig. \ref{fig2}a). Thus $\tau_{0H}$ and correspondingly $R_{H}$ change sign at
half -- filling (Fig. \ref{fig2}f). Because of electron -- hole symmetry
$\tau_{0}(n)=\tau_{0}(0.5+n)$ ($n\leq 0.5$) and derivative
$\frac{d\tau_{0}(n)}{dn}$ (this means also $S$) also can change the sign only
at half -- filling (cf. Fig. \ref{fig2}e).

If electron -- hole symmetry is broken ($t'/t=-0.4$) the region quadratic
electron dispersion close to $\Gamma$ -- point of the Brillouin zone becomes
very narrow, while the region of quadratic hole dispersion close to
M -- point becomes much wider (cf. Fig. \ref{fig2}b).
Correspondingly the region where the Fermi surface is electron -- like pocket
around $\Gamma$ -- point is limited to small fillings $n<0.3$, while in a wide
region of fillings $n\geq 0.3$ the Fermi surface is hole -- like pocket around
M -- point of the Brillouin zone (Fig. \ref{fig2}d).
Thus the sign change of $\tau_{0H}$ and correspondingly of $R_H$ is observed
for fillings $n\approx 0.4$ below half -- filling (Fig. \ref{fig2}e). More so,
the positive values of $\tau_{0H}(n)$ for $n<0.4$ is significantly smaller
than the absolute value of $\tau_{0H}(n)$ for $n>0.4$, so that the ``smearing''
due to the derivative of distribution function
$\left(-\frac{\partial f}{\partial \varepsilon}\right)$
in Eq. (\ref{Gxy1}) at high temperatures leads to $R_H$ changing its sign
at much lower fillings (Fig. \ref{fig1}d). In particular, for $T/8t=0.06$
$R_H$ changes sign at $n\approx 0.2$.

On the contrary, $\tau_{0}(n)$ for $t'/t=-0.4$ (Fig. \ref{fig2}e) has a
maximum at $n$ larger than half -- filling, so that $\frac{d\tau_{0}(n)}{dn}$
(and correspondingly $S$) changes its sign for $n\approx 0.66$.
Naturally the behavior of $\tau_{0}(n)$ is more or less  symmetric around its
maximum, and ``smearing'' due to the derivative of distribution function
in Eq. (\ref{S1}) close to this maximum leads only to small changes of $S$,
so that the filling $n\approx 0.66$, where thermopower changes its sign is
only weakly dependent on the temperature growth (cf. Fig. \ref{fig1}b).

Thus with the growth of $|t'|$ the band -- filling, corresponding to the
sign change of $R_H$, moves to the region of $n$ below half -- filling of the
band, while the band -- filling, corresponding to sign change of $S$,
more farther into the region of $n>0.5$. As a result, the growth of
$|t'|$ close to half -- filling leads to formation of a wider region of
band -- fillings, where Hall coefficient and thermopower have different signs,
and the growth of the temperature makes this region even wider.
Below we shall see that this tendency is observed also in systems with
systems with strong electronic correlations.

\section{Thermopower in strongly correlated metal and doped Mott insulator.}

Before going to the results of our DMFT calculations we shall present an
elementary qualitative analysis along the lines of Refs. \cite{KKKS1,KKKS2}.
It is more or less obvious that deep in the Mott insulator state with well --
defined upper and lower Hubbard bands under the hole doping both thermopower and
Hall coefficient are in fact determined by filling of the lower Hubbard band
(the upper band is much upper in energy and is practically empty).
Under this situation in the model with electron -- hole symmetry (in two --
dimensional case this corresponds to a spectrum with  $t'=0$) we can estimate
the band -- filling, corresponding to sign change of thermopower, in very
simple way. We shall consider only the paramagnetic phase with
$n_{\uparrow}=n_{\downarrow}=n$, so that in the following $n$ denotes electron
density per single spin projection and total electron density is $2n$.
In the following we consider only hole -- doping of Mott insulator and the
number of these ``holes'' is given by $p=1-2n$.

In the case of electron -- hole symmetry it is natural to suppose that both
thermopower and Hall coefficient change sign close to half -- filling of the
lower Hubbard band $n_0\approx 1/2$. Consider the states with ``upper'' spin
projection, then the full number of available states in the lower Hubbard band
id $1-n_{\downarrow}=1-n$.
Then the band -- filling is determined by
$n=n_{\uparrow}=n_0(1-n)\approx 1/2(1-n)$. Thus for the filling at which
both thermopower and Hall effect we obtain the estimate $n_c\approx 1/3$,
with corresponding ``hole'' concentration $p_c=1-2n_c\approx 1/3$
\cite{KKKS1,KKKS2}.


\begin{figure}
\includegraphics[clip=true,width=0.48\textwidth]{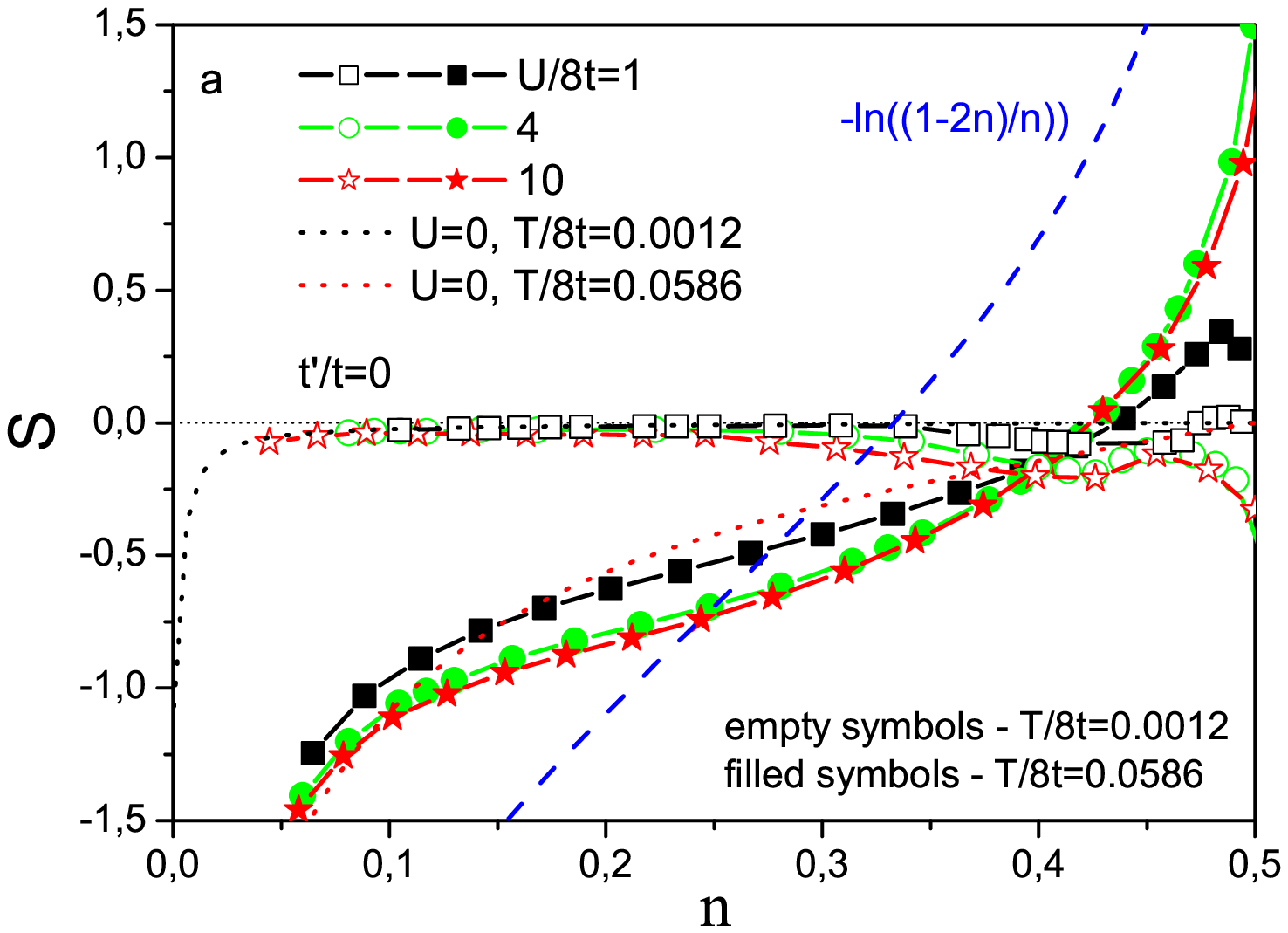}
\includegraphics[clip=true,width=0.48\textwidth]{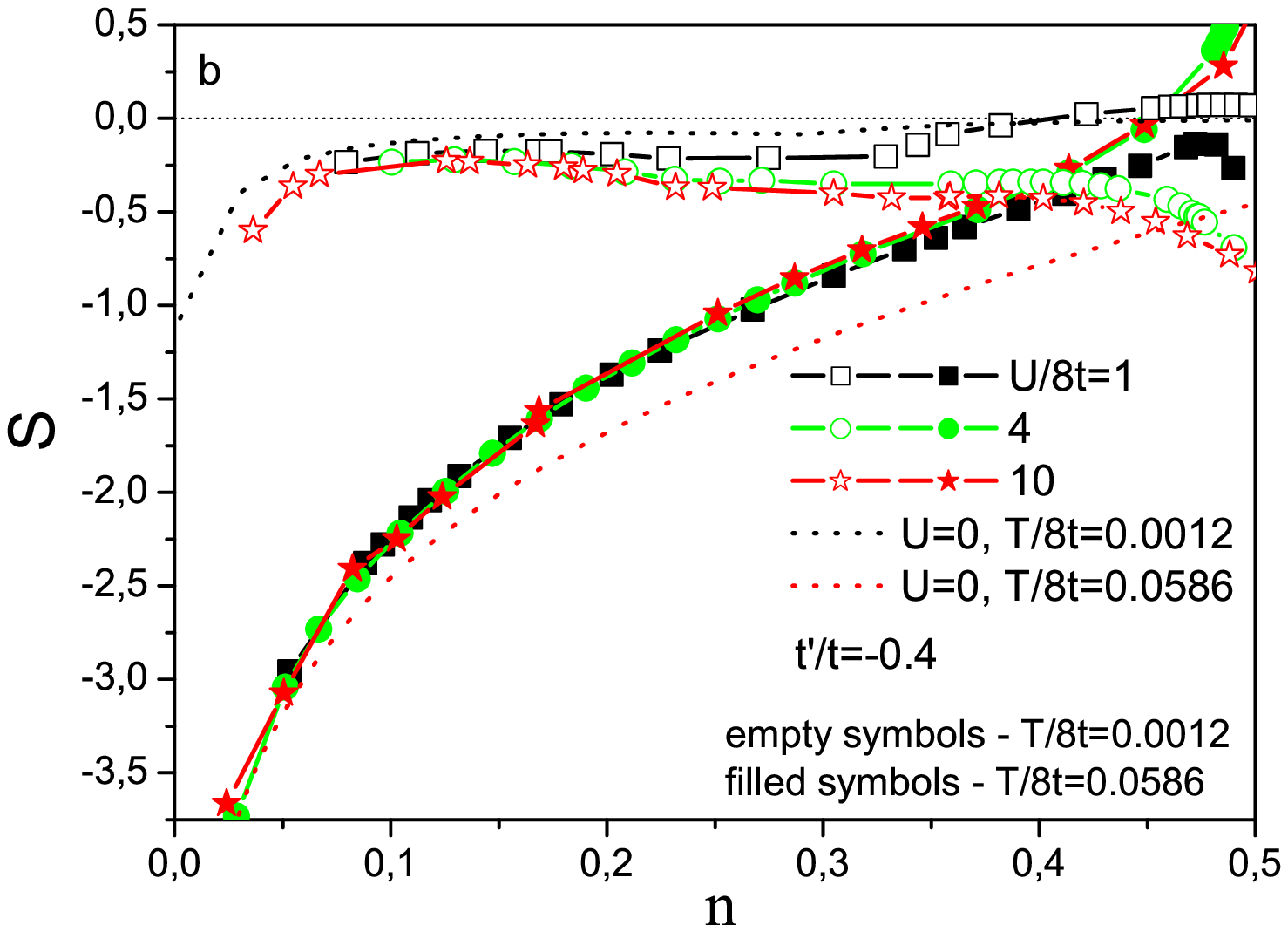}
\caption{Dependence of thermopower for the case of $t'=0$ (a) and $t'/t=-0.4$ (b)
on band -- filling for different strength of electron correlations $U$ in the
low -- temperature regime (empty symbols) and in high -- temperature regime
(filled symbols).}
\label{fig3}
\end{figure}


In general case situation is obviously more complicated.
In strongly correlated systems both thermopower and Hall effect (as well as
other electronic properties) are significantly dependent on temperature.
At low temperatures in these systems considered in DMFT approximation along
with lower and upper Hubbard bands an additional narrow quaiparticle band
forms close to the Fermi level leading to the so -- called quasiparticle peak
in the density of states \cite{pruschke,georges96,Vollh10}. In doped Mott
insulator this peak is placed near the upper edge of the lower Hubbard band.
Thus, at low temperatures both thermopower and Hall coefficient are mainly
determined by filling of this quasiparticle band. At high enough temperature
(of the order or larger than the width of quasiparticle peak) this peak is
smeared and thermopower is completely determined by the filling of the lower
Hubbard band. Correspondingly, during the analysis of thermopower, as well as
Hall effect \cite{KKKS1,KKKS2}, it is necessary to consider two different
temperature regimes.

\begin{figure}
\includegraphics[clip=true,width=0.48\textwidth]{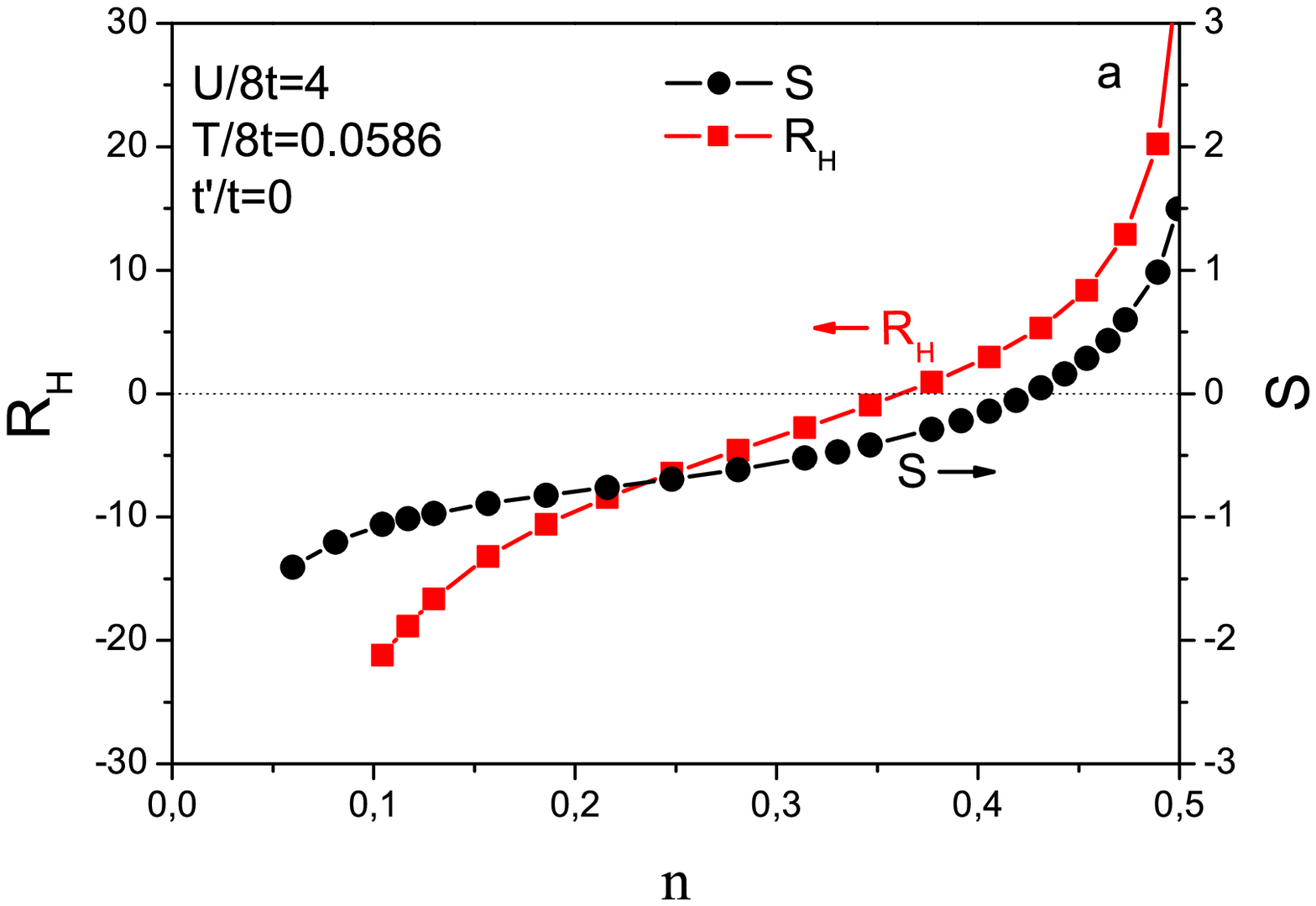}
\includegraphics[clip=true,width=0.48\textwidth]{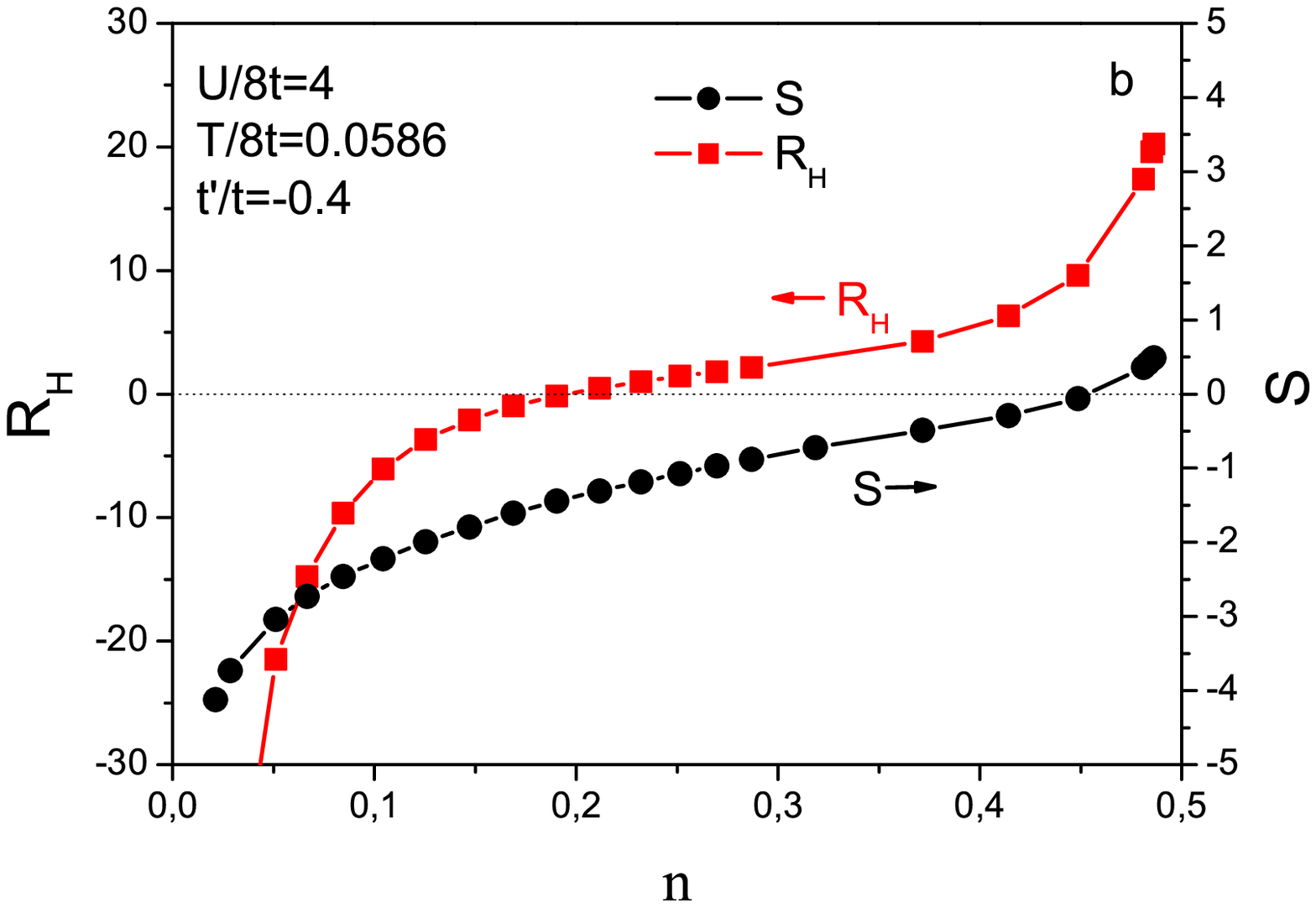}
\caption{Comparison of high -- temperature dependencies of thermopower and Hall
coefficient \cite{KKKS1,KKKS2} on band -- filling for $t'=0$ (a) and
$t'/t=-0.4$ (b)}
\label{fig3_1}
\end{figure}

In Fig. \ref{fig3} we show the dependence of thermopower on band -- filling,
both for strongly correlated metal ($U/8t=1$) and doped Mott insulator
($U/8t=4,10$). We see that in high -- temperature regime (curves with filled
symbols) thermopower, even in the model with complete electron -- hole
symmetry with $t'=0$ (Fig. \ref{fig3}a), in doped Mott insulator changes sign
at filling much closer to half -- filling ($n=0.42$), than follows from our
qualitative estimate. Note that Hall coefficient \cite{KKKS1,KKKS2} in this case
changes its sign at $n_{H}\sim 0.36$ (cf. Fig. \ref{fig3_1}a), which is quite
close to our qualitative estimate.
For comparison in Fig. \ref{fig3}a blue dashed line shows an exact result
for thermopower \cite{Beni1974}:
\begin{equation}
S=-\frac{k_B}{e}ln\frac{2p}{1-p}=-\frac{k_B}{e}ln\frac{1-2n}{n}
\label{Beni}
\end{equation}
for the limit of $U\gg t$. It can be seen that in this atomic limit thermopower
changes sign at $n=1/3$ in complete accordance with our estimate.
It is possible that this significant difference is due to our use of DMFT
approximation. Cluster DMFT \cite{Phillips2010} and Monte -- Carlo calculations
\cite{S_MC} give thermopower values for $U\gg t$ much closer to an exact result
of Eq. (\ref{Beni}). However the main deficiency of these approaches is the
possibility to perform calculations only in the region of high enough
temperatures $T\sim t$.

For noticeable breaking of electron -- hole symmetry our qualitative estimate
becomes invalid both for the Hall coefficient \cite{KKKS1,KKKS2} and
thermopower (cf. Fig. \ref{fig3}b). Note that in accordance with our analysis
of the case of $U=0$ given above the deviation from electron -- hole symmetry
leads to decreasing values of the filling, corresponding to change of the sign of
Hall coefficient \cite{KKKS1,KKKS2}, and increasing values of the filling,
corresponding the sign change of thermopower (cf. Fig. \ref{fig3_1}b).
Thus, in strongly correlated systems breaking of electron -- hole systems
leads to the appearance of rather wide region of band -- fillings, where Hall
coefficient and thermopower have different signs.
To compare high -- temperature behavior of thermopower and Hall coefficient
\cite{KKKS1,KKKS2} in hole doped Mott dielectric we show corresponding
band -- filling dependencies in Fig. \ref{fig3_1} for $t'/t=0; -0.4$.

In the low -- temperature limit (curves with unfilled symbols in Fig. \ref{fig3})
the presence of quasiparticle peak leads to the absence of sign change of
thermopower in doped Mott insulator, which remains negative (electron -- like)
in the whole region of hole dopings.
Note that Hall coefficient at low temperatures becomes positive (hole -- like)
only in a narrow region close to half -- filling, i.e. at very small hole
doping \cite{KKKS1,KKKS2}. In low -- temperature regime the width and the
amplitude of quasiparticle peak depend both on band -- filling and temperature.
Significant dependence of quasiparticle peak on band -- filling in low --
temperature regime leads to the appearance of rather wide region of band --
fillings, where thermopower decreases with the growth of $n$ (cf. Fig. \ref{fig3}).
Such anomalies on filling, related to quasiparticle peak, are stronger for
thermopower than for Hall coefficient \cite{KKKS1,KKKS2}.

Smearing and disappearance of quasiparticle peak may be not only due to the
growth of temperature, but also due to disordering \cite{GDMFT,dis_hubb_2008},
as well as due to pseudogap fluctuations, which are totally neglected in the
local DMFT approach \cite{GDMFT,DMFT+S}.
Thus, the region of applicability of simple estimates and qualitative behavior
of thermopower, given above in the framework of DMFT for high -- temperature
regime, in reality may be much wider.

In general case calculation of disorder scattering effects (more so pseudogap
fluctuations) upon thermopower is rather complicated problem. As a simple
estimate below we present the results of calculations using Eqs. (\ref{S1})
and (\ref{tau1}), where we use the values of spectral density
$A({\bf p}\varepsilon)$ for disordered Hubbard model, obtained within
DMFT+$\Sigma$ approach \cite{GDMFT,DMFT+S}. Disorder parameter $\Delta$ denotes
here an effective scattering rate of electrons by the random field of impurities
(in self -- consistent Born approximation). It is clear that such approach,
taking into account only the influence of disorder in the spectral density, is
oversimplified, but it seems reasonable for qualitative analysis.


\begin{figure}
\includegraphics[clip=true,width=0.48\textwidth]{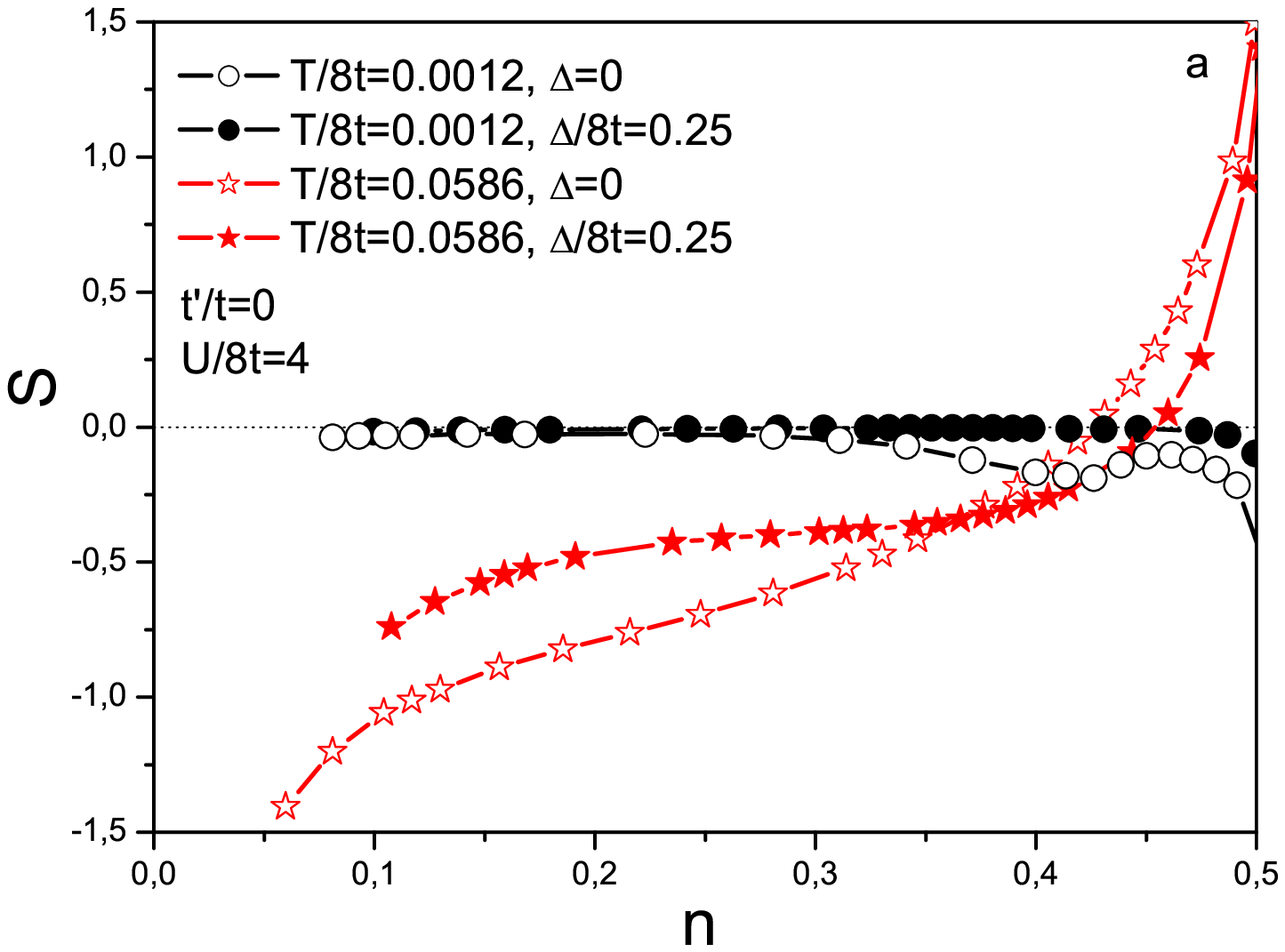}
\includegraphics[clip=true,width=0.48\textwidth]{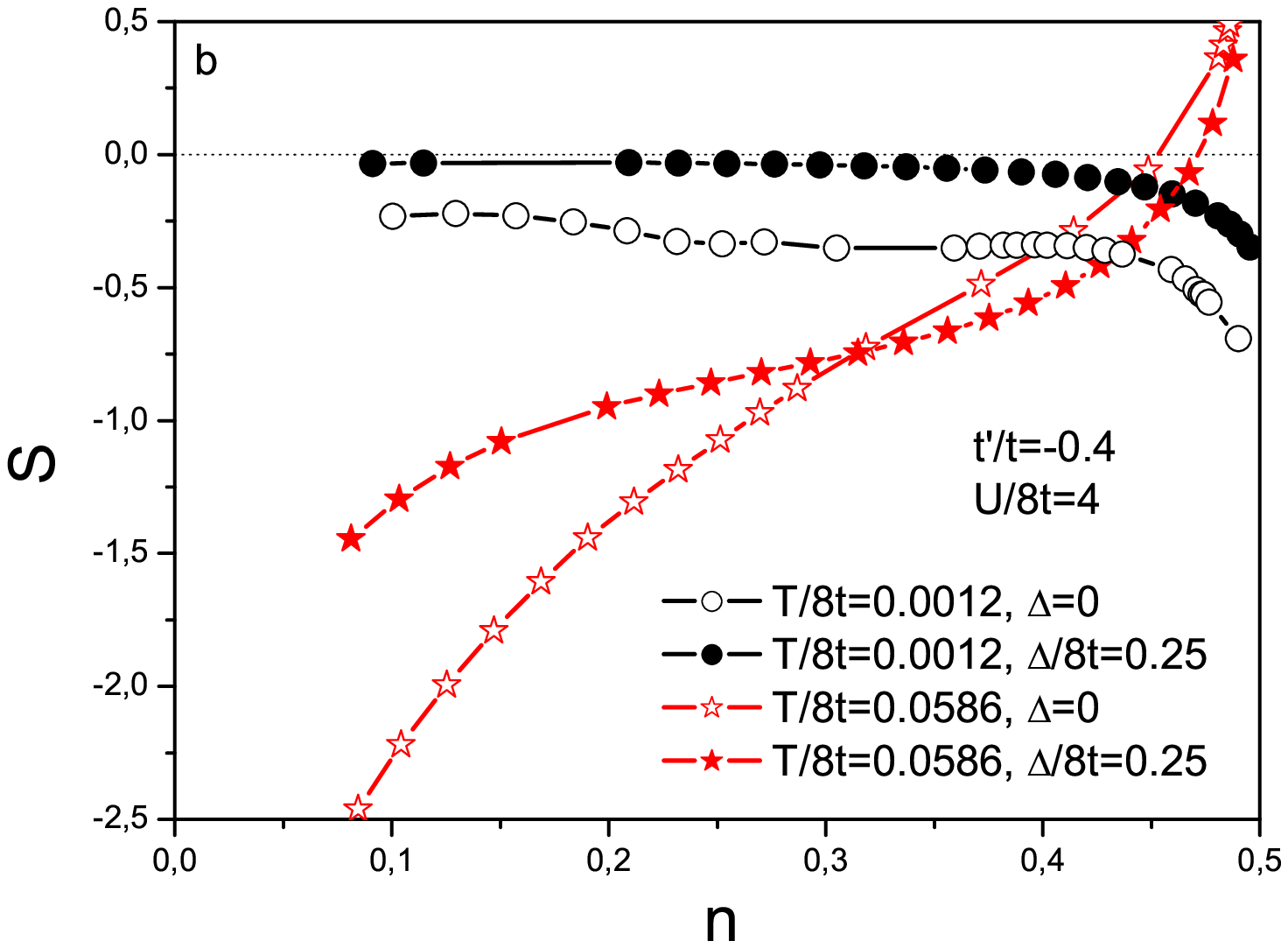}
\caption{Dependence of thermopower on bandfilling in the presence of impurity
scattering ($\Delta/8t=0.25$, filled symbols) and in its absence
($\Delta=0$, empty symbols) for $t'=0$ (a) and $t'/t=-0.4$ (b)}
\label{fig4}
\end{figure}


In Fig. \ref{fig4} we compare dependencies of thermopower on band -- filling
in the absence of disorder (empty symbols) and for the case impurity scattering
with $\Delta/8t=0.25$ in Mott insulator with $U/8t=4$.
While in high -- temperature regime disorder only weakly affects the Hall
coefficient \cite{KKKS1,KKKS2} for different values of $t'/t$, for the
thermopower disorder influence, even in this regime, is rather noticeable.
Disorder leads to the decrease of thermopower for fillings close to half --
filling (low hole doping), leading to the decrease of hole doping,
corresponding to sign change of $S$. At high enough hole doping< when thermopower
is negative, the growth of disorder leads to rather significant decrease
of the absolute value of $S$. In low -- temperature regime, where thermopower
is negative at any hole dopings, increasing disorder leads to a strong decrease
of the absolute value of thermopower.

\begin{figure}
\includegraphics[clip=true,width=0.48\textwidth]{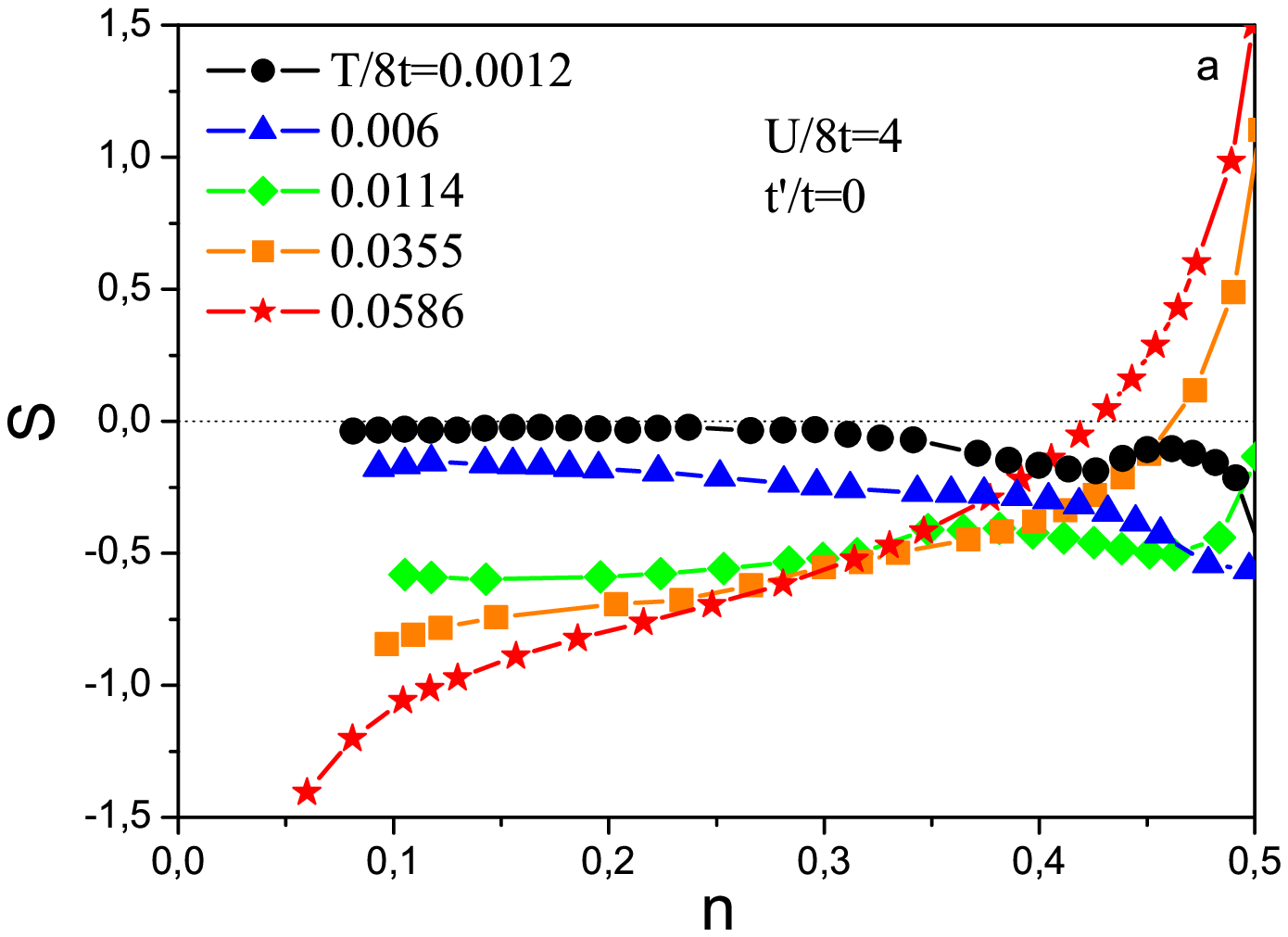}
\includegraphics[clip=true,width=0.48\textwidth]{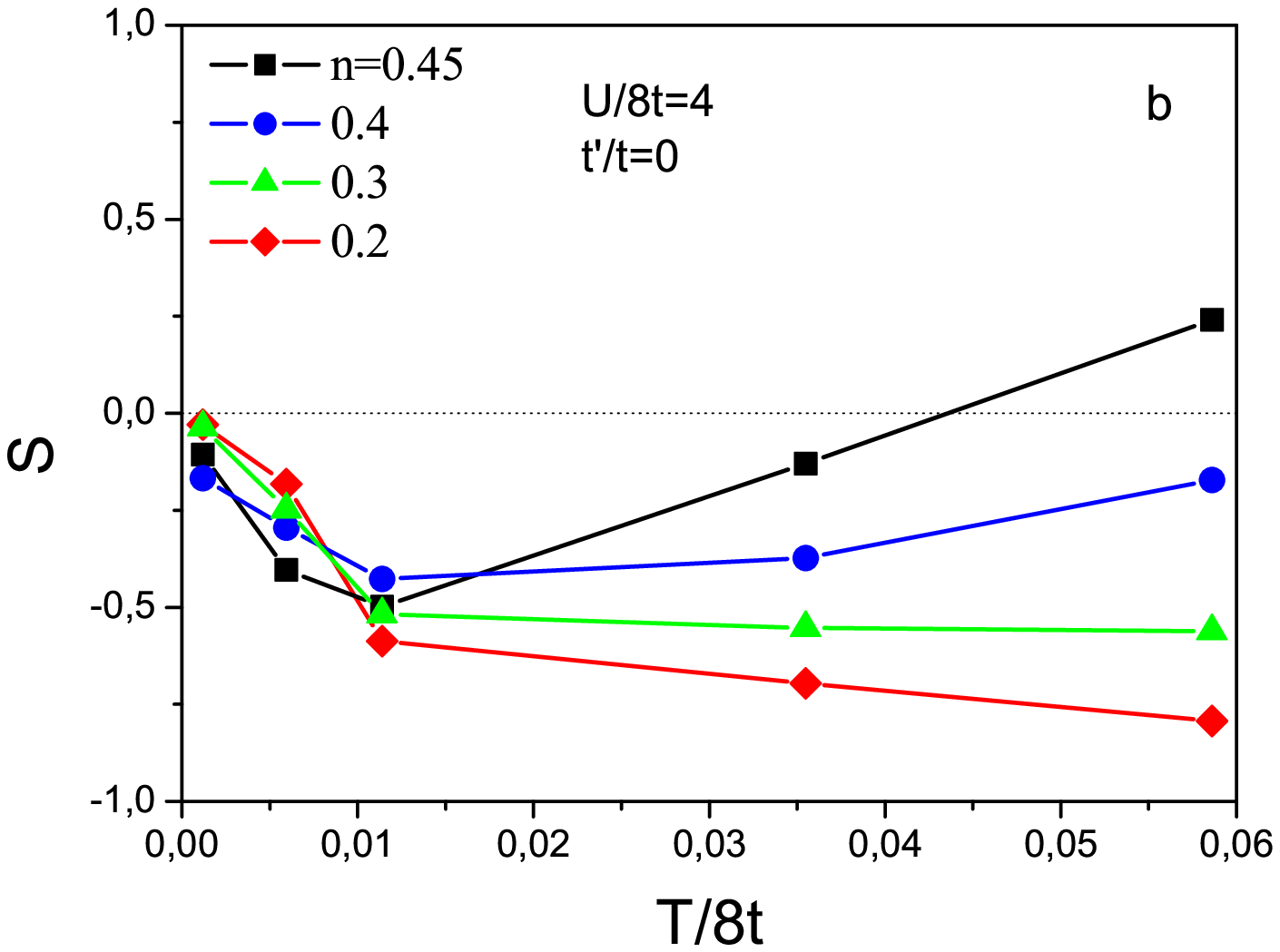}
\includegraphics[clip=true,width=0.48\textwidth]{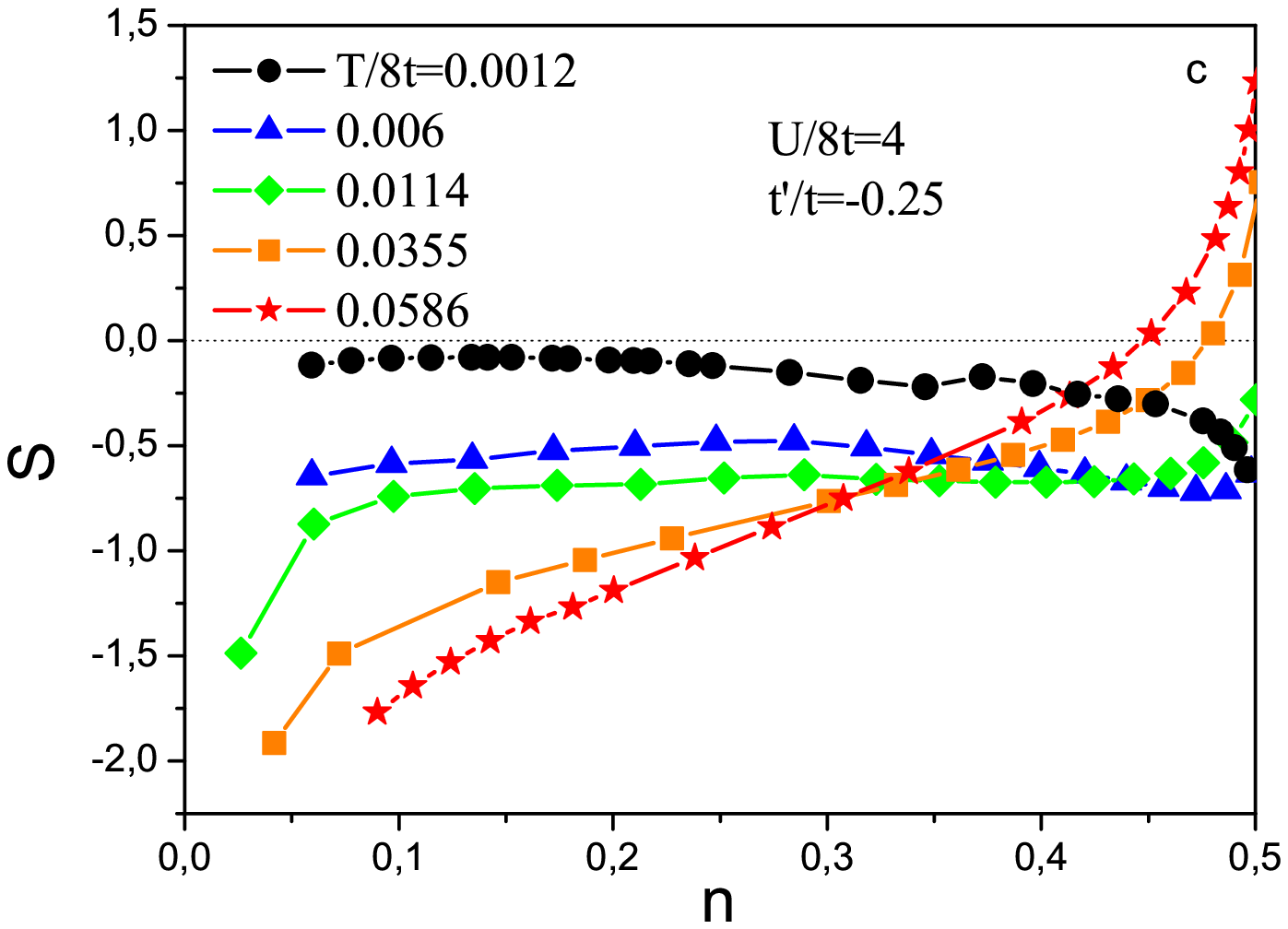}
\includegraphics[clip=true,width=0.48\textwidth]{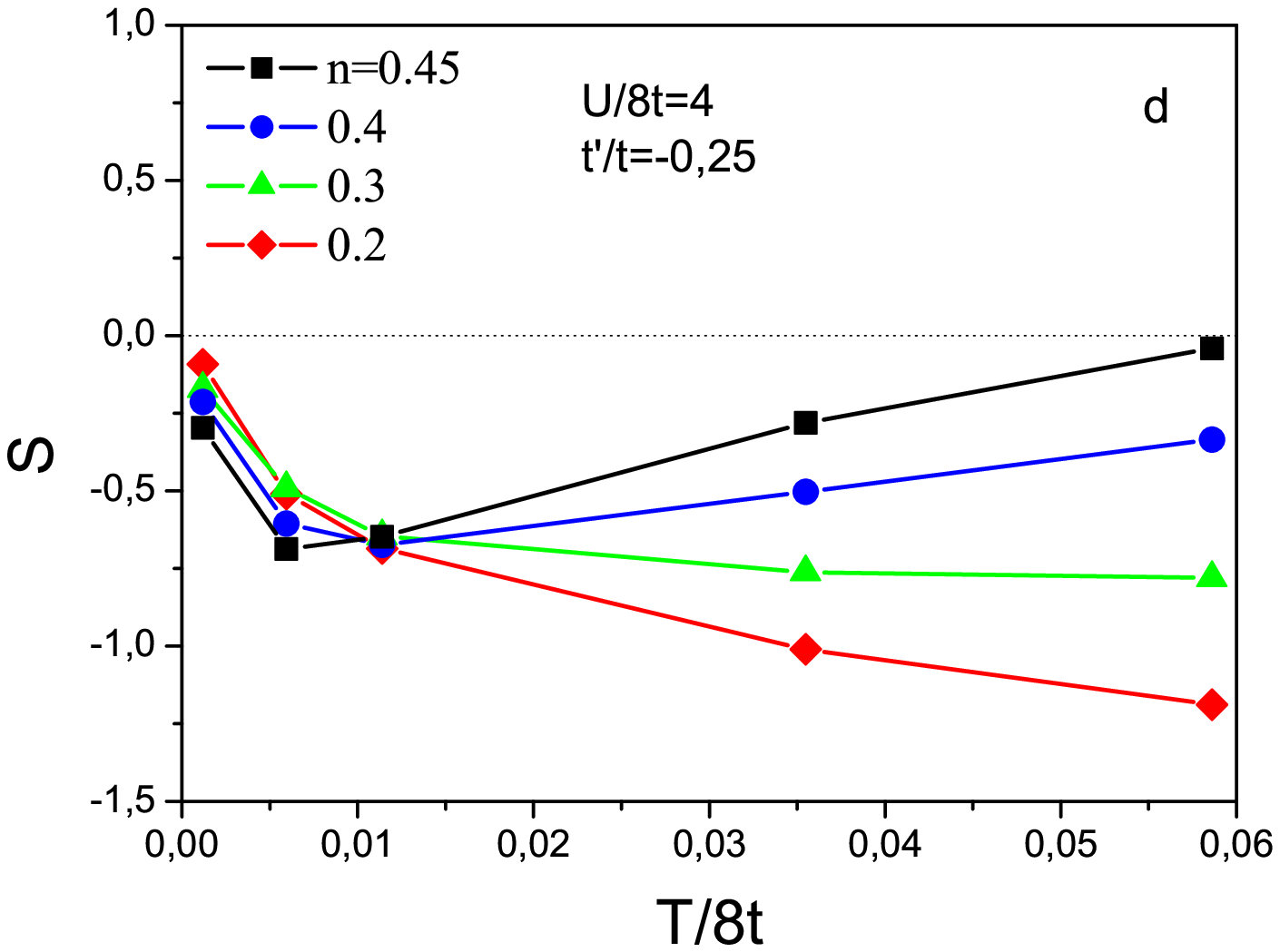}
\includegraphics[clip=true,width=0.48\textwidth]{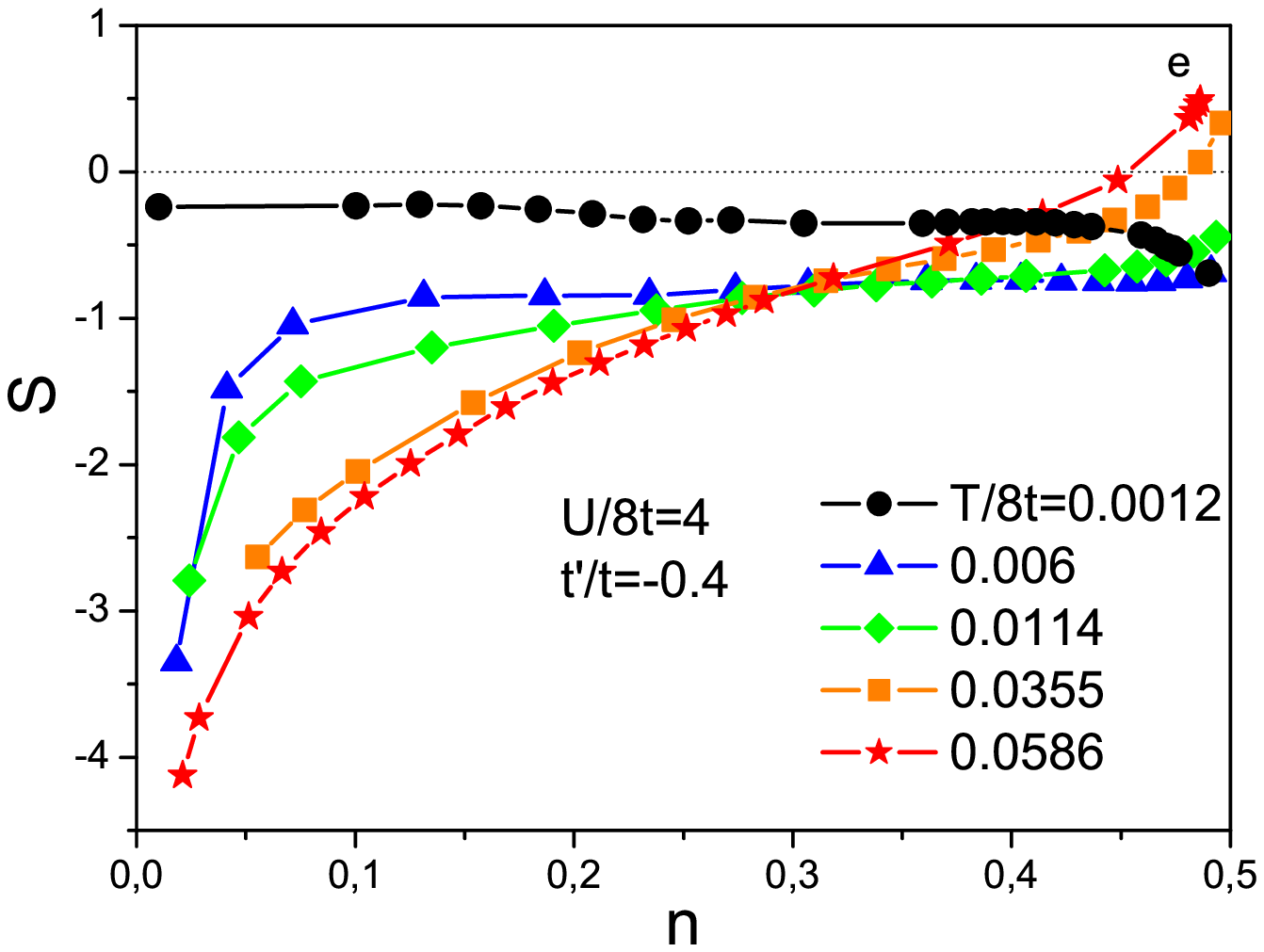}
\includegraphics[clip=true,width=0.48\textwidth]{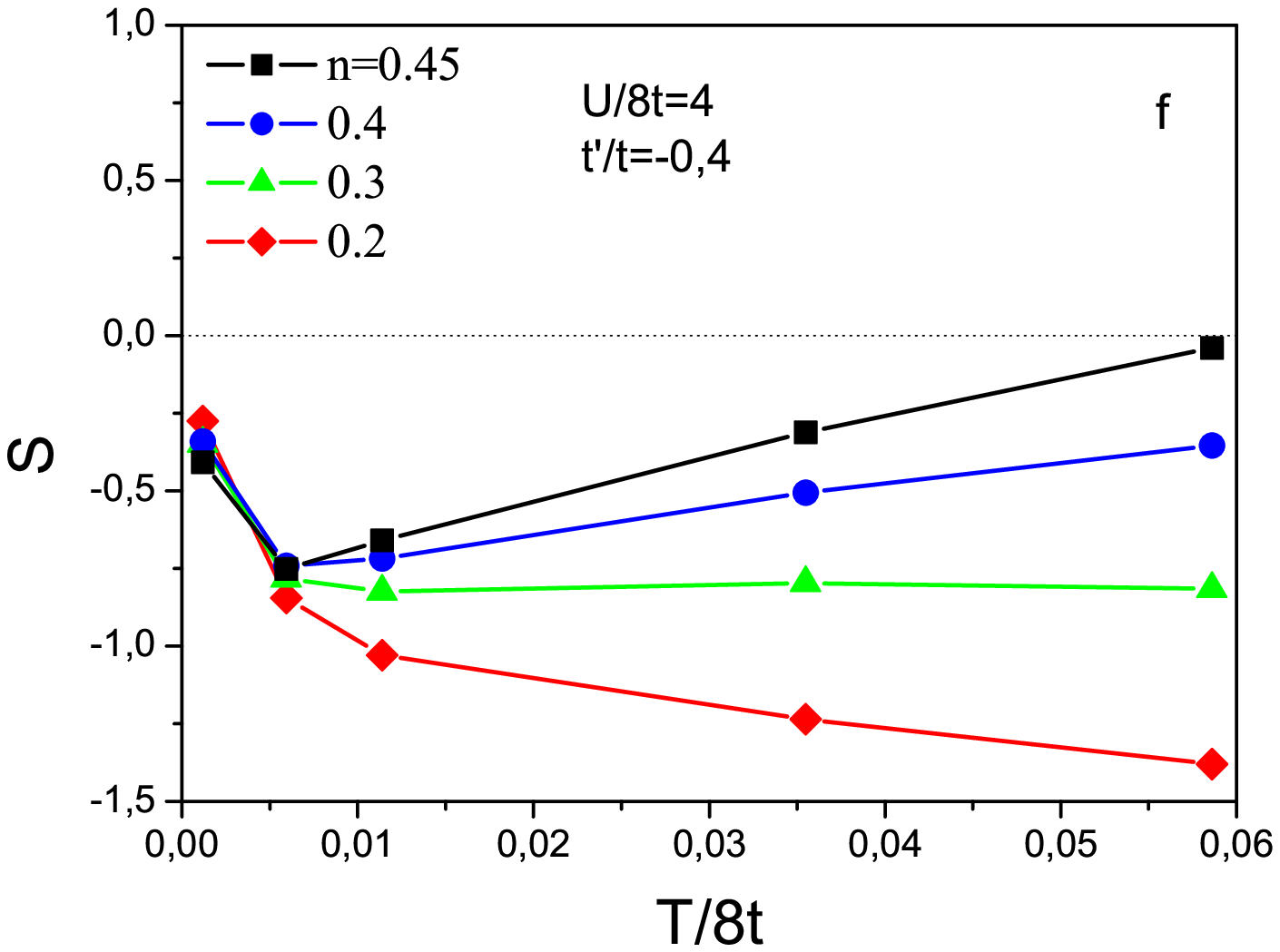}
\caption{Dependence of thermopower on band -- filling for different temperatures
(a,c,e) and temperature dependence of $S$ for different band -- fillings (b,d,f).}
\label{fig5}
\end{figure}

In Fig. \ref{fig5} we show the dependencies of thermopower on band -- filling
and temperature in the case of Mott insulator with $U/8t=4$ for different
models of electron spectrum, for the case of complete electron -- hole
symmetry $t'=0$ and for $t'/t=-0.25$ and $t'/t=-0.4$, typical correspondingly
for cuprates like LSCO and YBCO. On the filling dependence of $S$ we observe
smooth evolution from low -- temperature to high -- temperature regime.
While at low temperatures for values of  $t'/t$, considered here, thermopower
is negative at all fillings, increasing of the temperature close to half --
filling leads to the appearance of the region of positive values of thermopower.
Hole doping at which $S$ changes its sign increases with temperature and
decreases with the growth of $|t'/t|$. It is necessary to mote, that for Hall
coefficient \cite{KKKS1,KKKS2} the growth of $|t'/t|$ leads to the decrease of
the region of hole dopings with positive values of Hall coefficient and  to the
increase of doping value, corresponding to the sign change of $R_H$.
Situation is here is in many respects similar to the case of absence of
correlations ($U=0$) -- the growth of $|t'/t|$  leads to the increase of
band -- filling, corresponding to sign change of thermopower and the decrease
of the filling, corresponding to sign change of $R_H$.

In Fig. \ref{fig5} b,d,f we show the temperature dependencies of thermopower for
different band -- fillings. In all cases we observe the significant dependence
of $S$ on temperature. For low temperatures thermopower is negative and its
absolute value increases with the growth of the temperature.
At high temperatures close to half -- filling (low hole doping) thermopower is
increasing with rising temperature changing sign of $S$, leading to positive
values of $S$ at high $T$. At small fillings (high hole doping) the absolute
value of $S$ at high temperatures continues to increase with the growth of
$T$, though slower than at low temperatures.

\begin{figure}
\includegraphics[clip=true,width=0.50\textwidth]{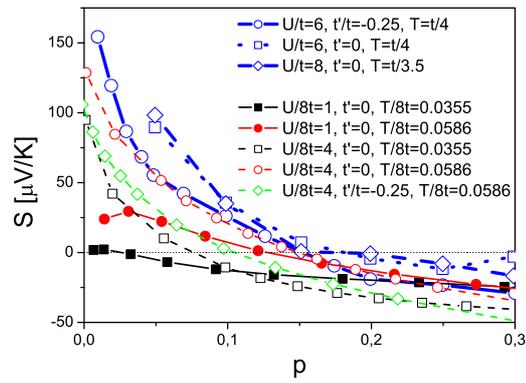}
\caption{Comparison of DMFT results and Monte -- Carlo calculations \cite{S_MC}
(thick blue lines) for the dependence of thermopower on hole doping $p$.}
\label{fig6}
\end{figure}

Recently quantum Monte -- Carlo calculations of thermopower were done for the
Hubbard model \cite{S_MC} at reasonable temperature $T=t/4$.
The results of these calculations nicely reproduce the experimental results
on thermopower for a number of hole -- doped cuprates
(cf. Fig. 1 in Ref. \cite{S_MC}).  In particular they reproduce the sign change
of thermopower at hole doping $p\approx 0.15$. In Fig. \ref{fig6} we compare
the results of our DMFT calculations with Monte -- Carlo results of
Ref. \cite{S_MC}. We can see that the results of our DMFT calculations of
thermopower at $U/8t=4$ are close to Monte -- Carlo results at
$U/8t\approx 1$, and correspondingly to the room temperature experimental data
for thermopower in cuprates. Note that in the framework of DMFT at $U/8t=1$
the system still remains (even at half -- filling) the strongly correlated metal,
while at $U/8t=4$ it becomes the doped Mott insulator, which is usually
considered as typical for cuprates.

\section{Estimating carrier concentration, thermopower and Hall coefficient from ARPES data.}

Measuring thermopower and Hall coefficient are among the major experimental
methods to determine the type and concentration of current carriers.
However, as was demonstrated above, especially under the conditions of broken
electron -- hole symmetry ($t'\neq 0$) a wide region of band -- fillings appear,
where thermopower and Hall effect produce evidence for for different type
of charge carriers ($p$--$n$ anomaly). As was shown in Refs. \cite{KKKS1,KKKS2}
in the interval of doping, where Hall effect changes its sign, Hall number
(the number of carriers formally determined from this effect) sharply increases
and can not be used to determine the real number of carriers.

Below we shall show, that a semiquantitative estimate of the number of charge
carriers, as well as of thermopower and Hall coefficient in doped Mott
insulator can be done using ARPES data and electron spectrum obtained within
the standard DFT calculations.

\begin{figure}
\includegraphics[clip=true,width=0.48\textwidth]{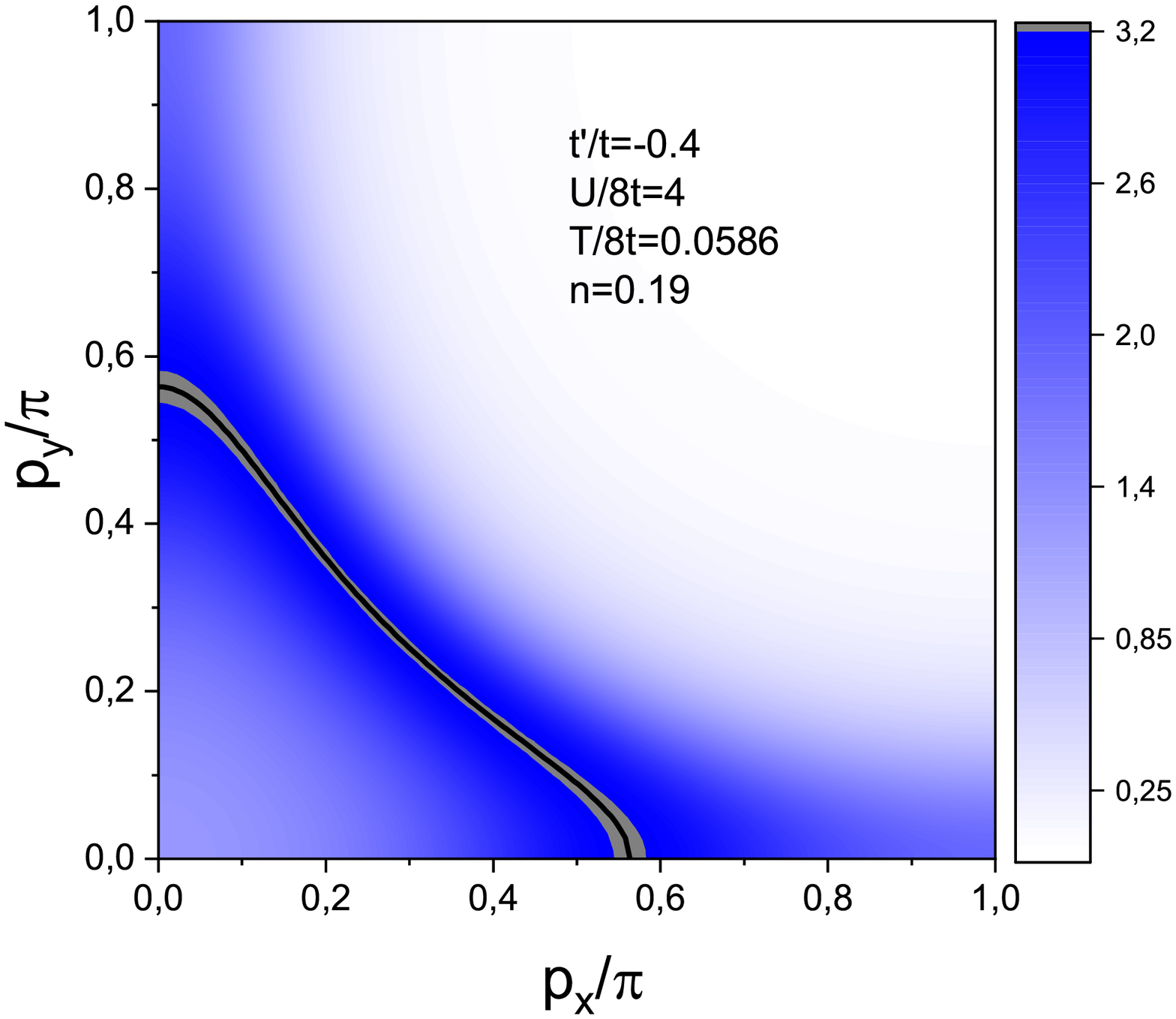}
\caption{Spectral density in the quarter of the Brillouin zone. Curve represents
the Fermi surface obtained from Eq. (\ref{SF1}).}
\label{fig7}
\end{figure}

In systems with strong electronic correlations, doped Mott insulator including,
Fermi -- liquid description still holds, Fermi surface is well defined , as well
as quasiparticles near it \cite{Kot}.
In Fig. \ref{fig7} we show the spectral density $A({\bf p},0)$ at the Fermi
level, obtained within DMFT in high -- temperature regime, when the temperature
is noticeably larger than the width of quasiparticle peak, but still much
lower than the Fermi energy. Band -- filling is taken to be equal to $n=0.19$,
at which, in the case of $t'/t=-0.4$, Hall coefficient changes its sign
\cite{KKKS1,KKKS2}. We see that spectral density is much ``smeared'', but it
still has a maximum at the Fermi surface, shown in Fig. \ref{fig7} by black
curve, obtained from equation:
\begin{equation}
\mu-\varepsilon({\bf p})-{\rm Re}\Sigma(0)=0,
\label{SF1}
\end{equation}
where $\mu$ is the chemical potential, determined from band -- filling within
DMFT, while $\Sigma(0)$ is local (DMFT) self -- energy at the Fermi level.

Let us introduce now $\mu_{eff}=\mu-{\rm Re}\Sigma(0)$, then instead of
Eq. (\ref{SF1}) we obtain an equation:
\begin{equation}
\mu_{eff}-\varepsilon({\bf p})=0,
\label{SF2}
\end{equation}
determining the Fermi surface like in the absence of electronic correlations
($U=0$), but with chemical potential $\mu_{eff}$. Obviously in general case
spectral density is much ``smeared'' around the Fermi surface, but its maximum
is still on it (Fig. \ref{fig7}). Thus for qualitative (but much simpler)
estimate of Hall coefficient and thermopower we can use expressions given by,
Eqs. (\ref{S5}) and (\ref{R_H3}), corresponding to $U=0$, but with
$\mu\rightarrow\mu_{eff}$. Naturally, the filling $n_{0}$, corresponding to
chemical potential $\mu_{eff}$ in the system with no correlations ($U=0$),
does not coincide with real band -- filling $n$ in correlated system.
It seems reasonable to assume that this $n_{0}$ corresponds to the
filling of the lower Hubbard band. Then from $n_{0}$ we can easily get the
total filling of the whole band $n$. The total number of states in the lower
Hubbard band is $1-n$, so that the total filling is $n=n_{0}(1-n)$ and we get:
\begin{equation}
n=\frac{n_{0}}{1+n_{0}}.
\label{n_n0}
\end{equation}

\begin{figure}
\includegraphics[clip=true,width=0.32\textwidth]{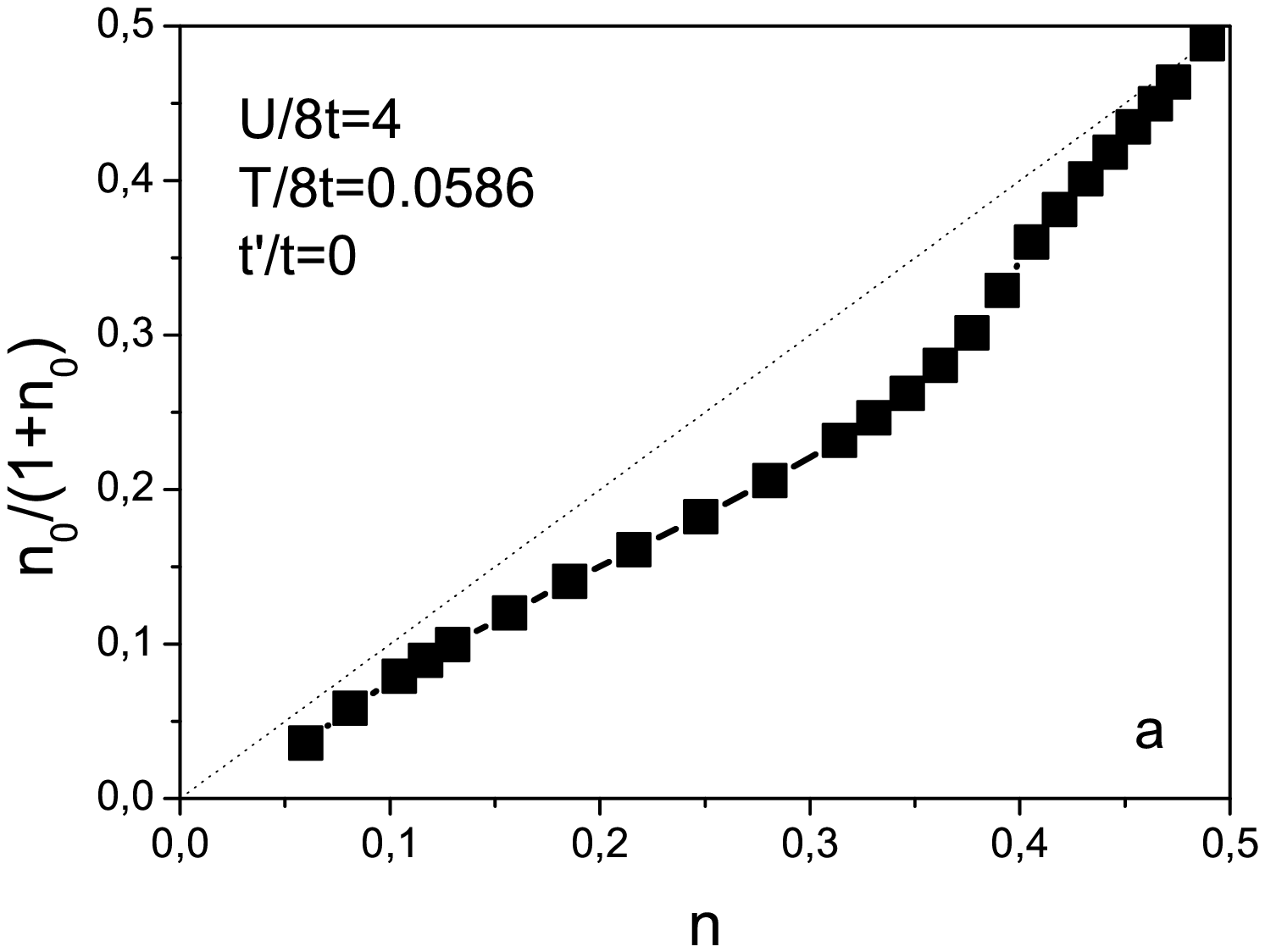}
\includegraphics[clip=true,width=0.32\textwidth]{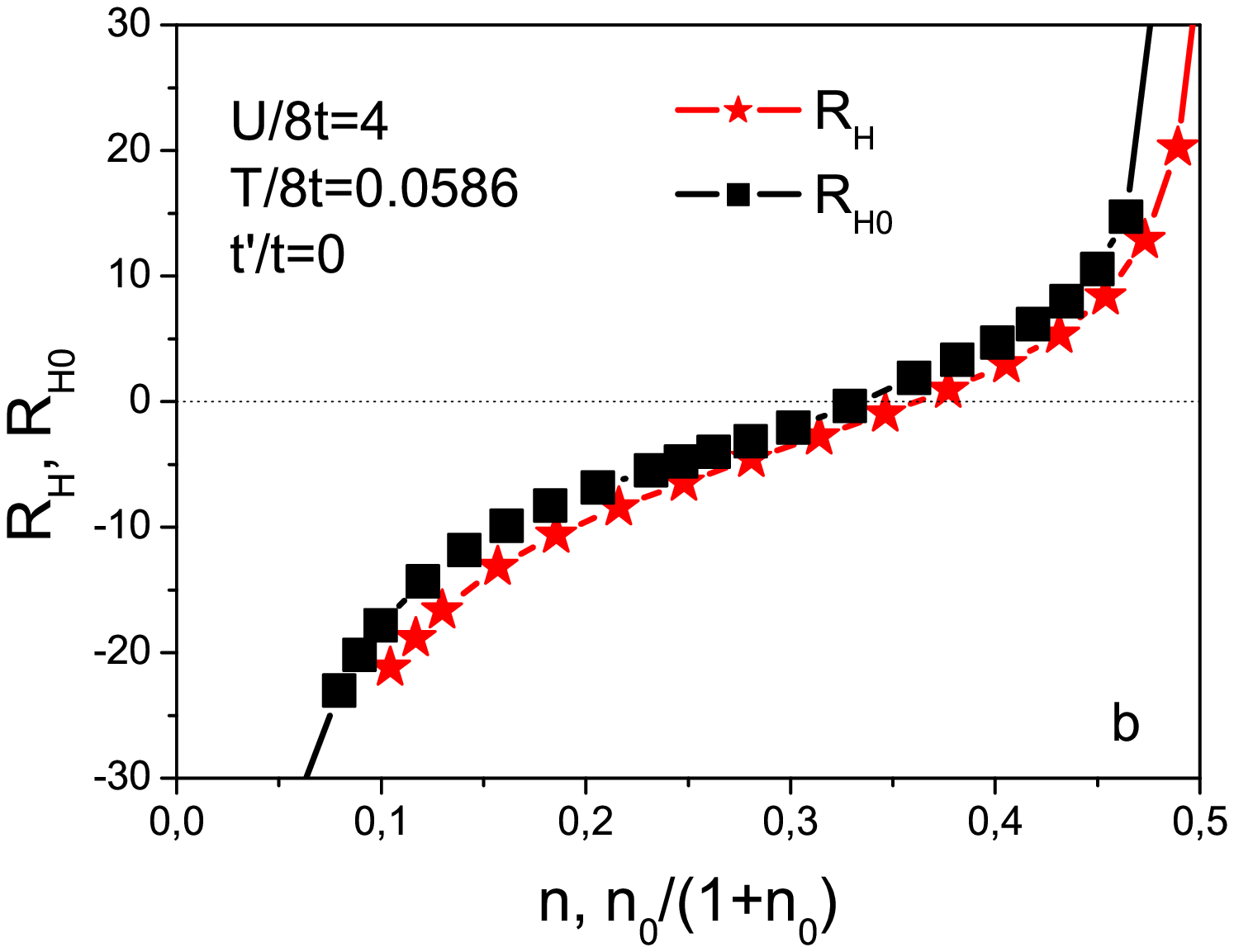}
\includegraphics[clip=true,width=0.32\textwidth]{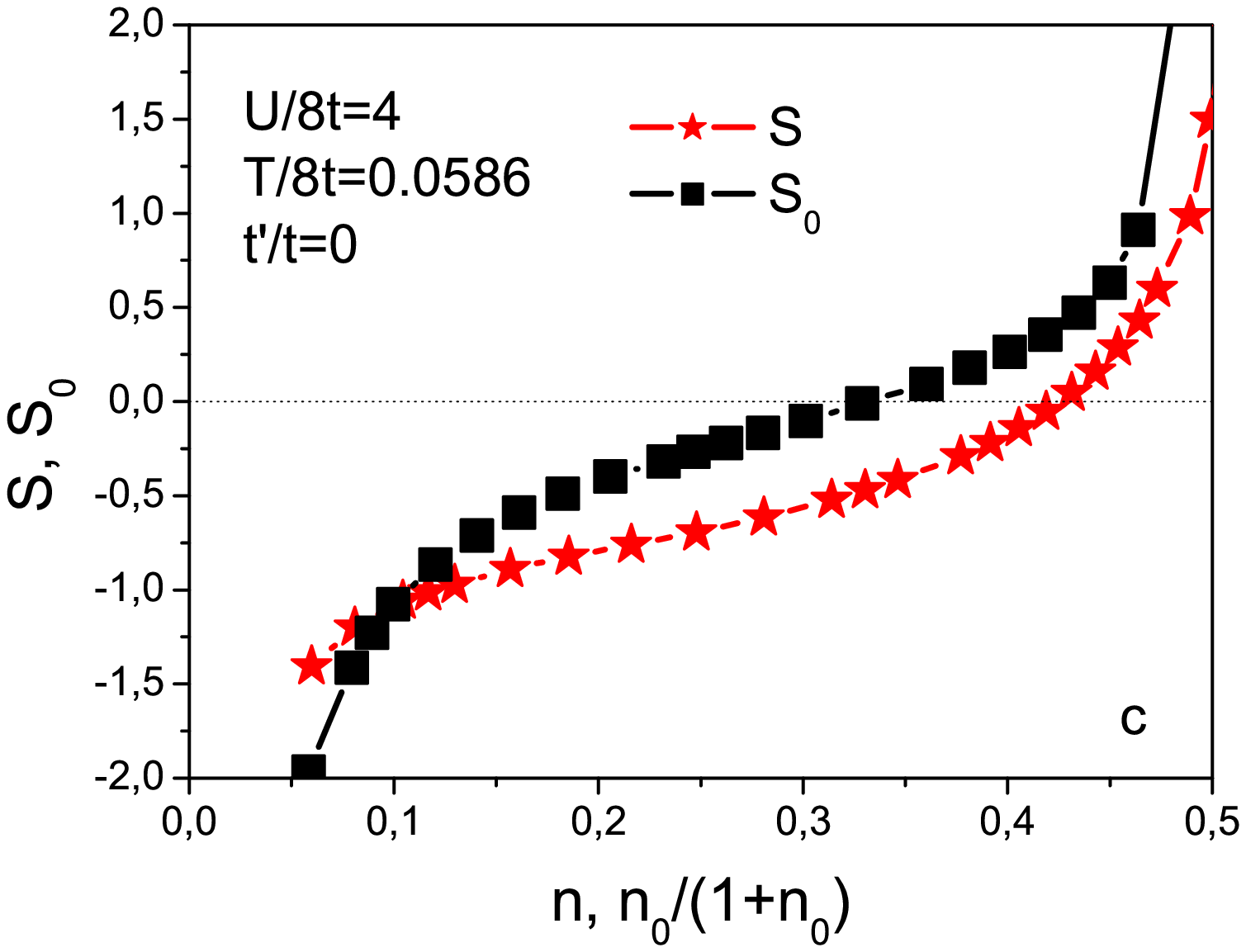}
\includegraphics[clip=true,width=0.32\textwidth]{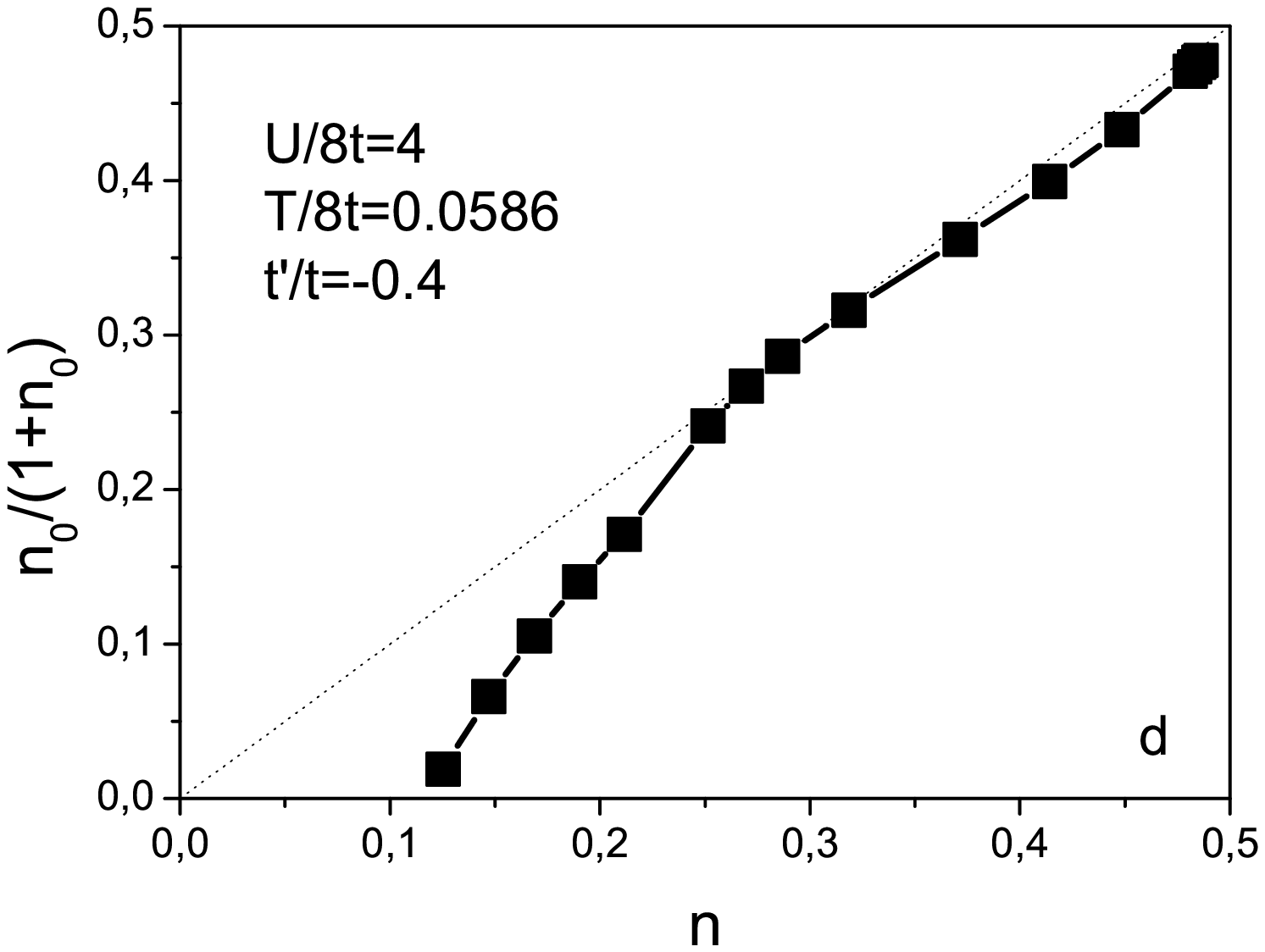}
\includegraphics[clip=true,width=0.32\textwidth]{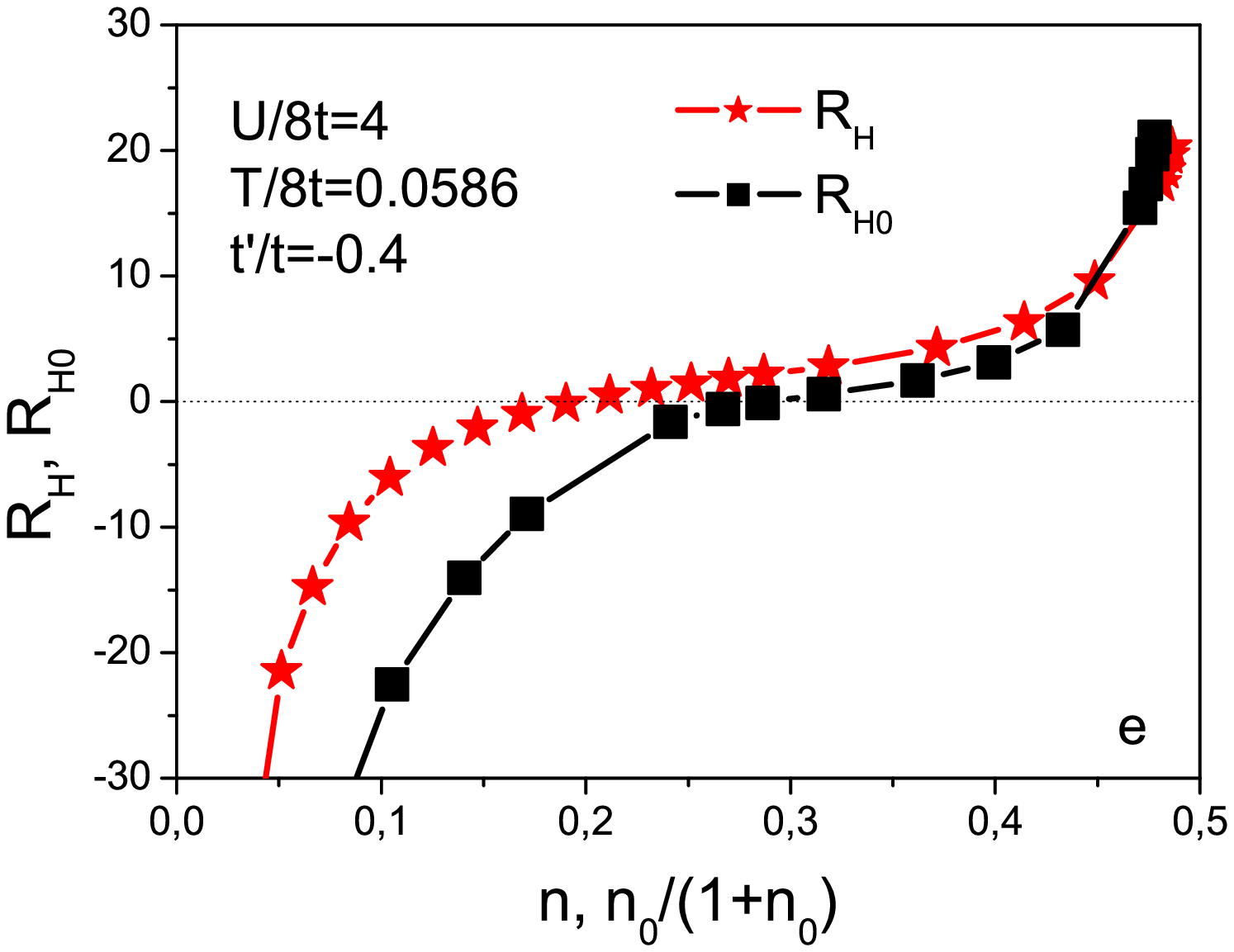}
\includegraphics[clip=true,width=0.32\textwidth]{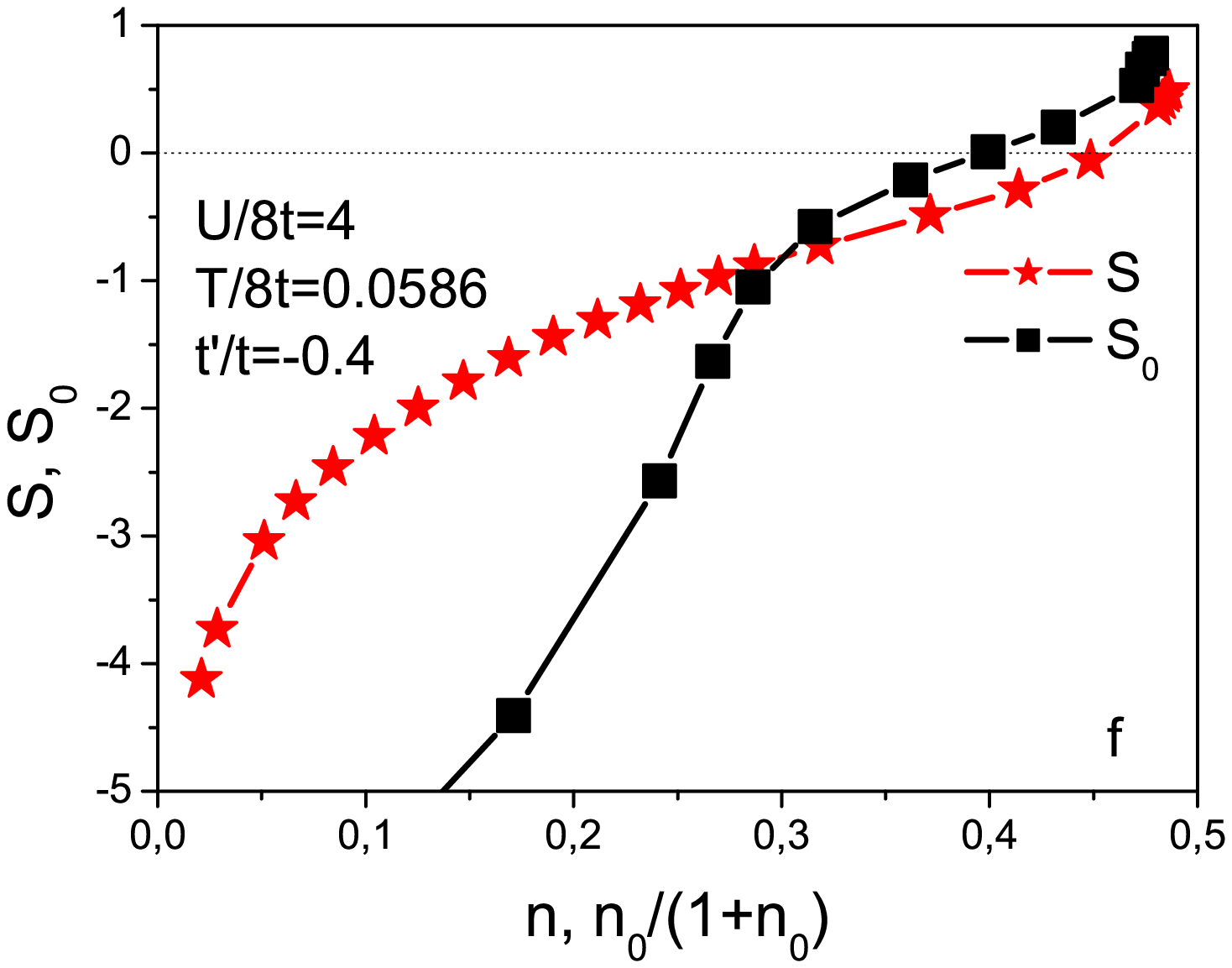}
\includegraphics[clip=true,width=0.32\textwidth]{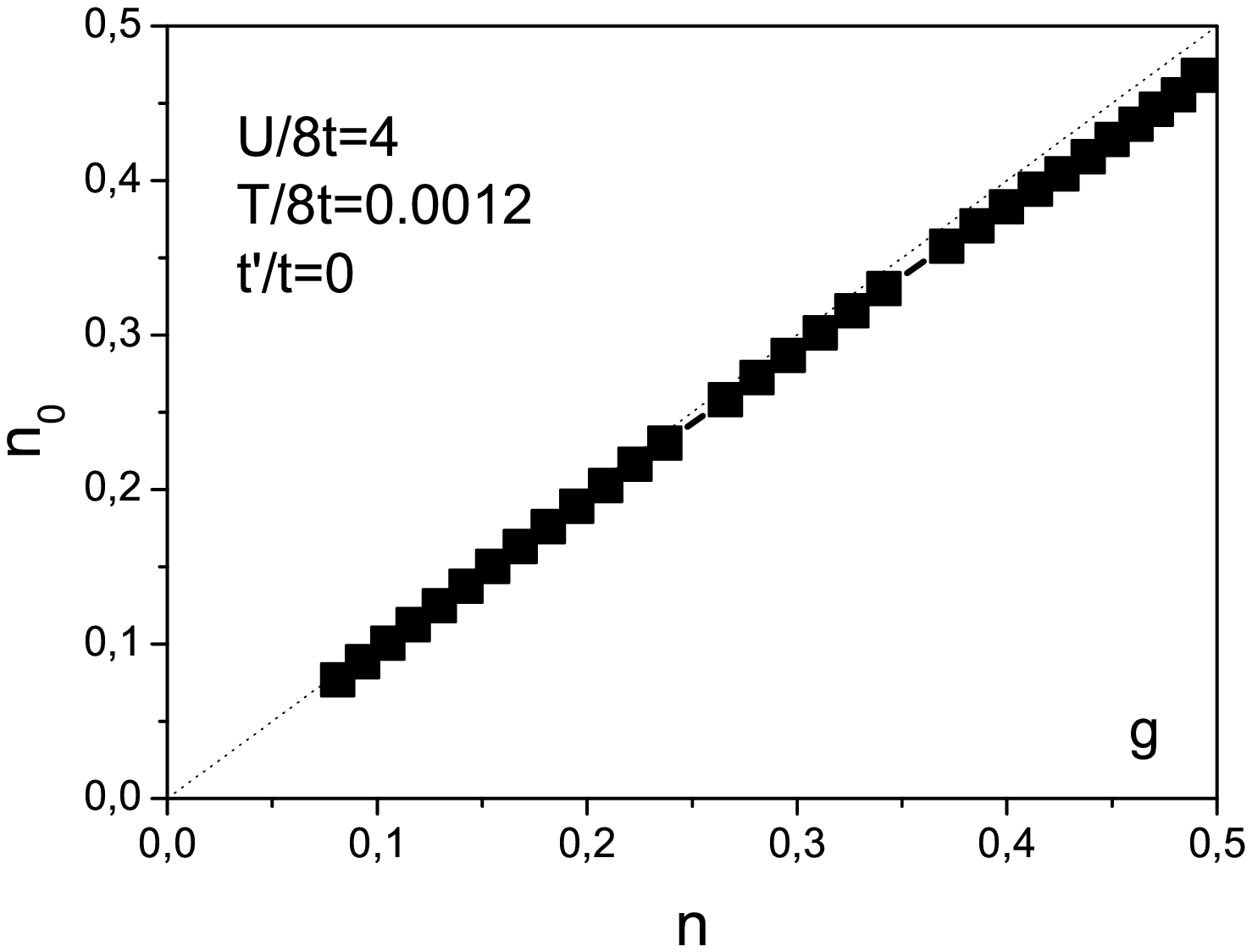}
\includegraphics[clip=true,width=0.32\textwidth]{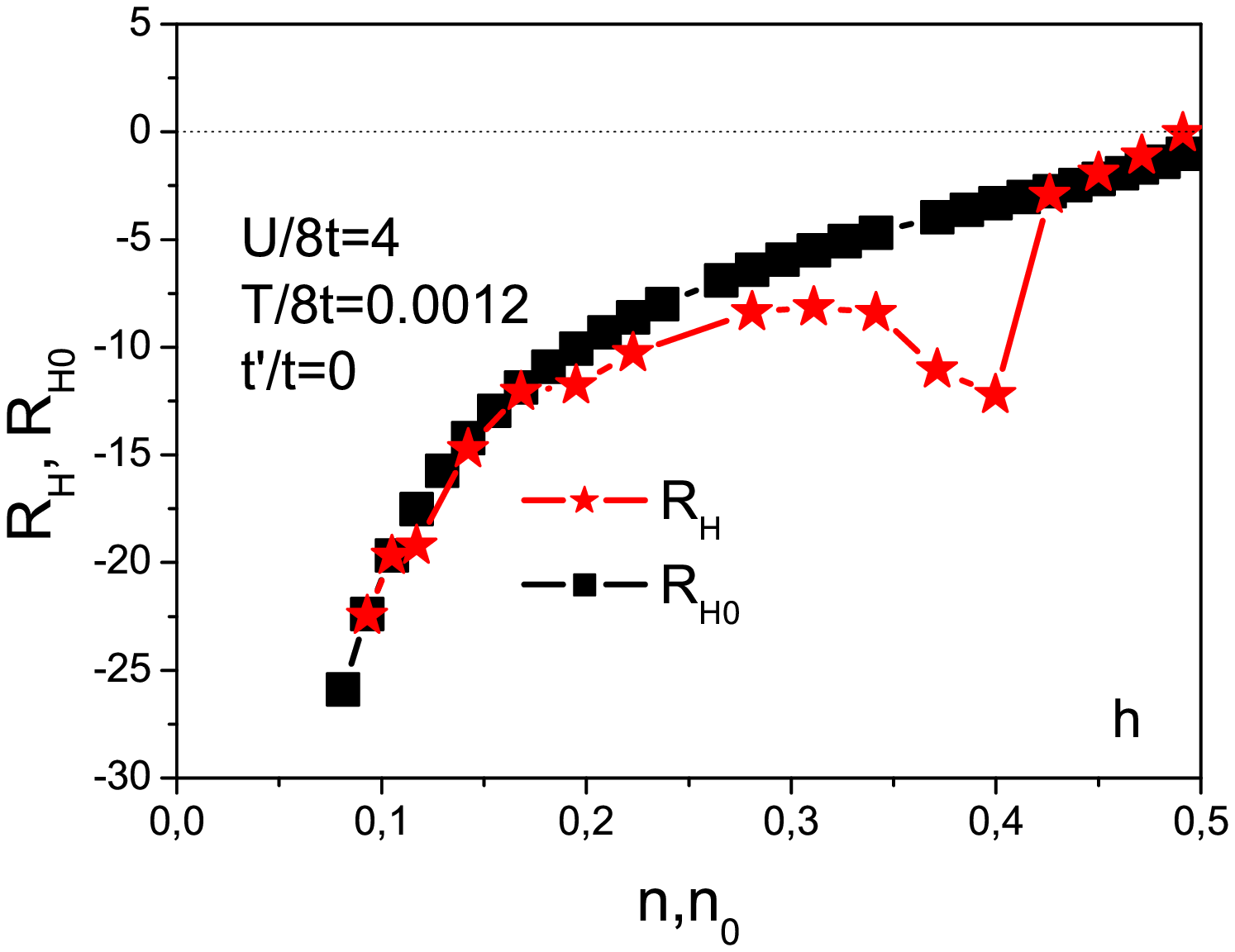}
\includegraphics[clip=true,width=0.32\textwidth]{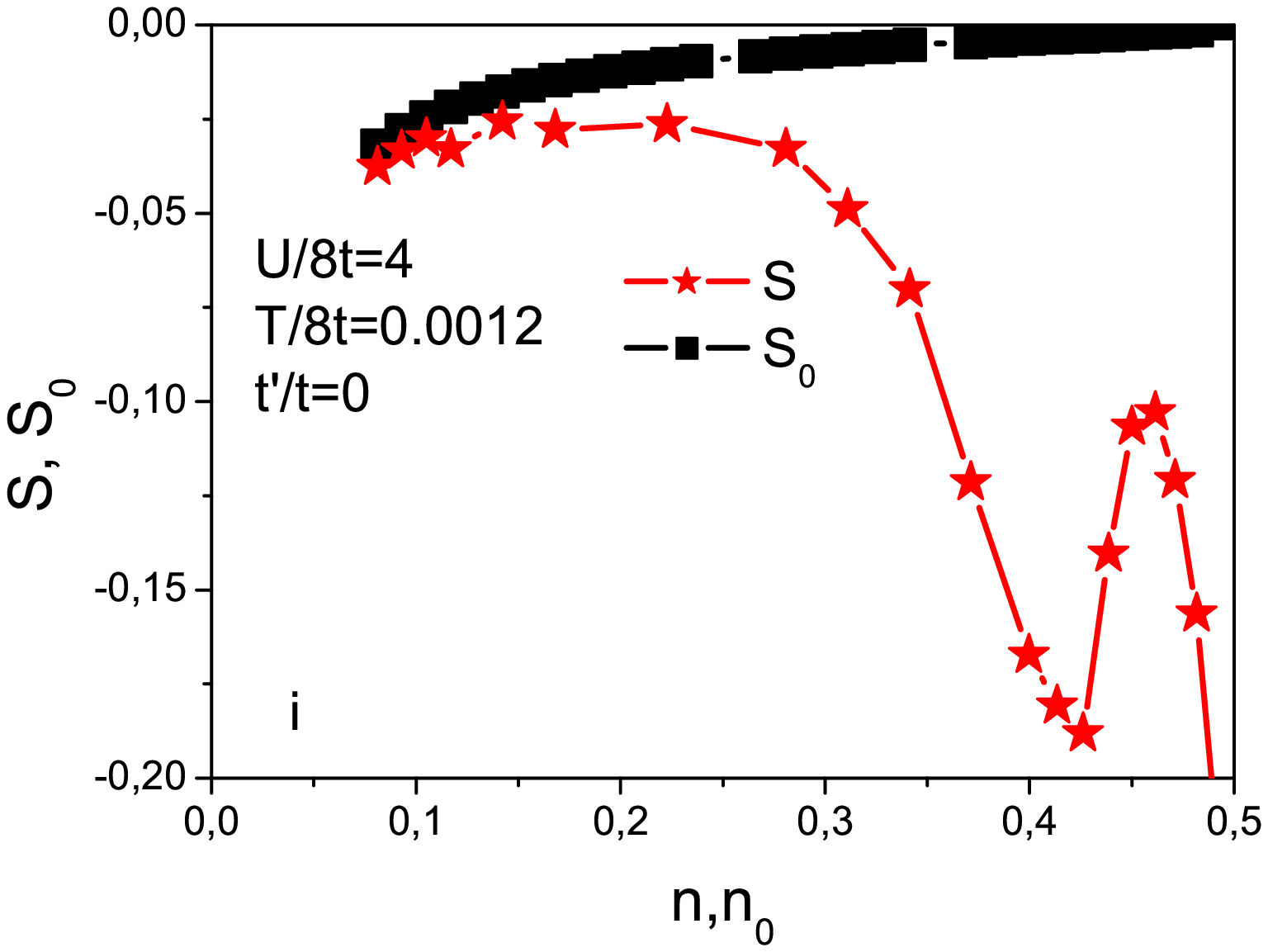}
\includegraphics[clip=true,width=0.32\textwidth]{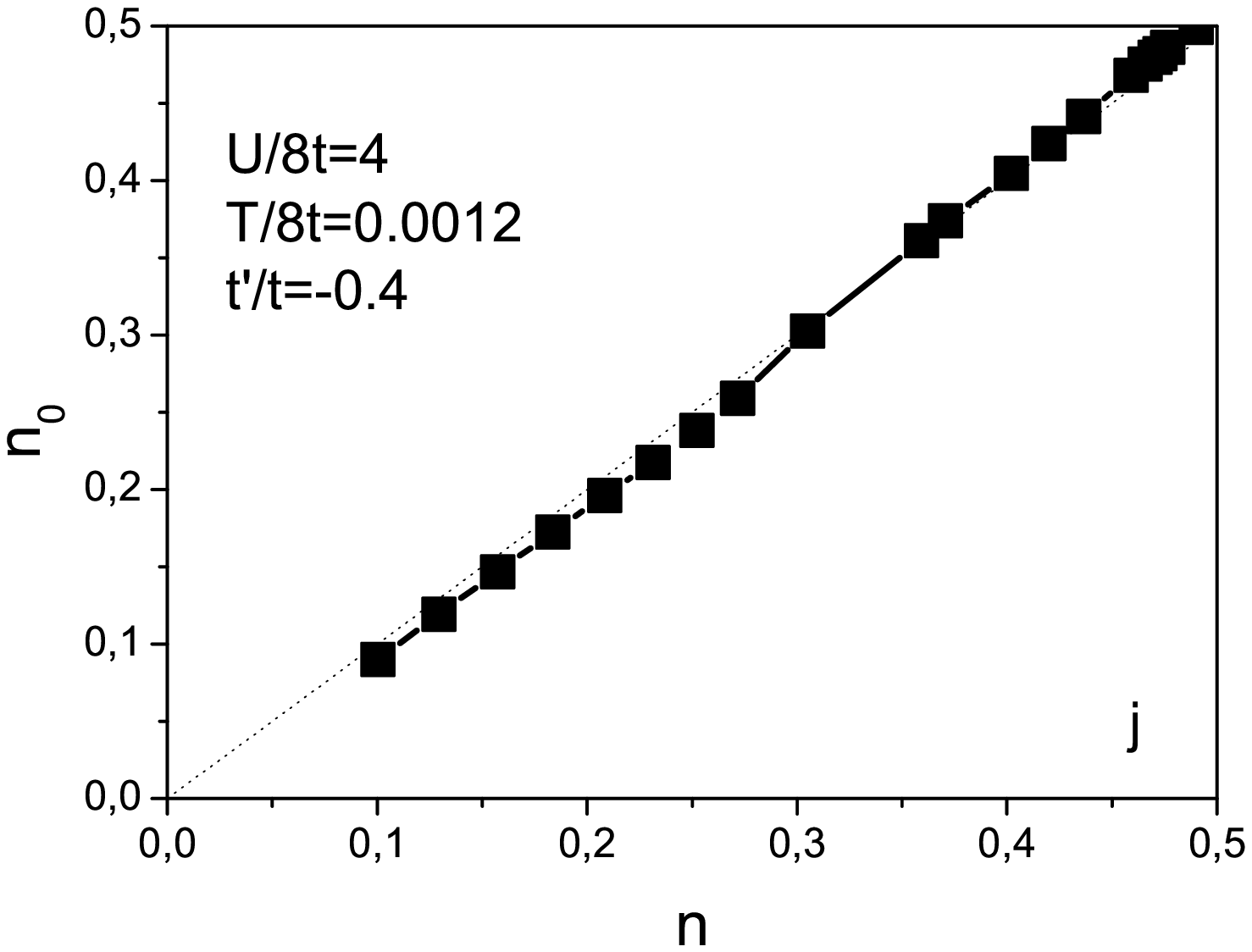}
\includegraphics[clip=true,width=0.32\textwidth]{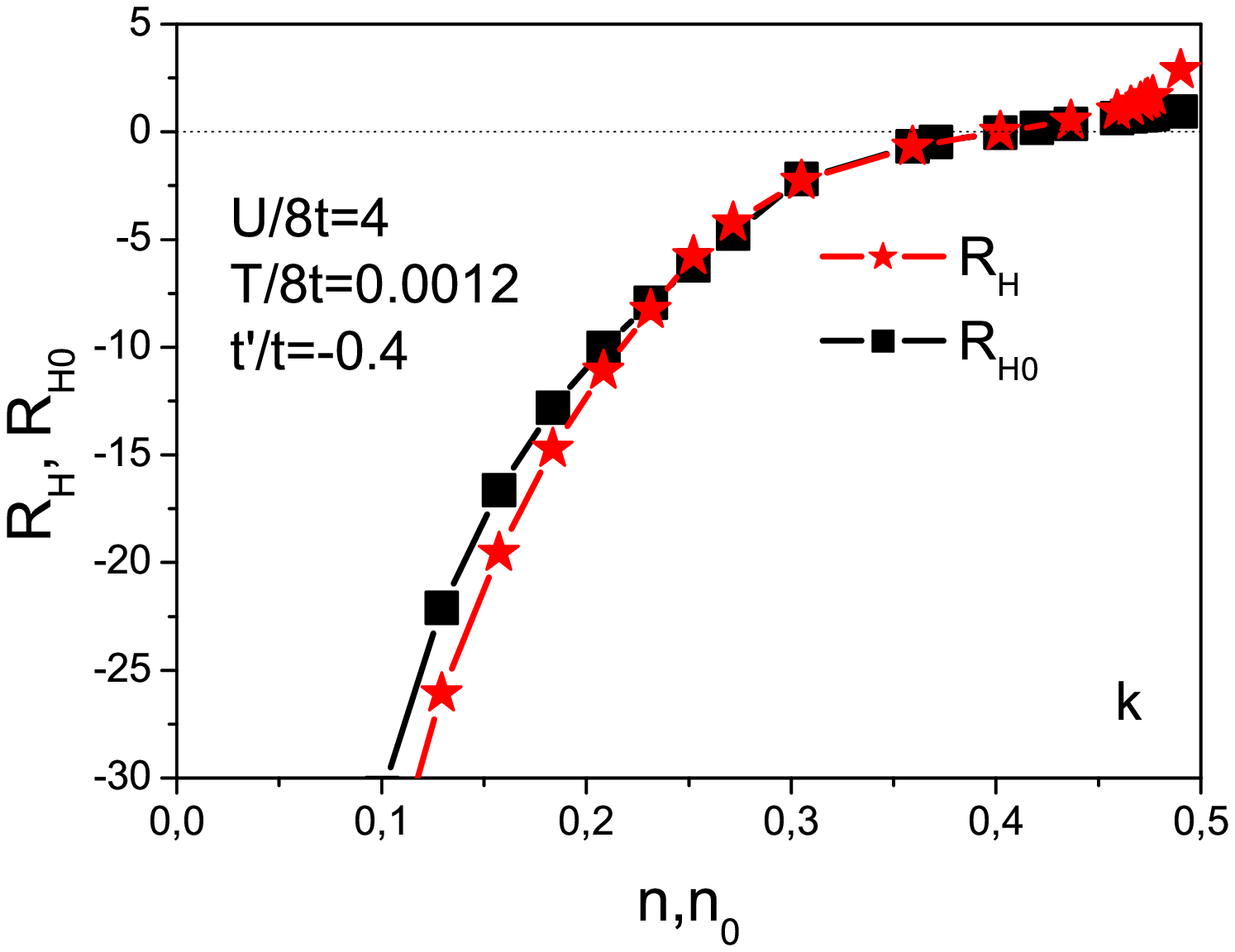}
\includegraphics[clip=true,width=0.32\textwidth]{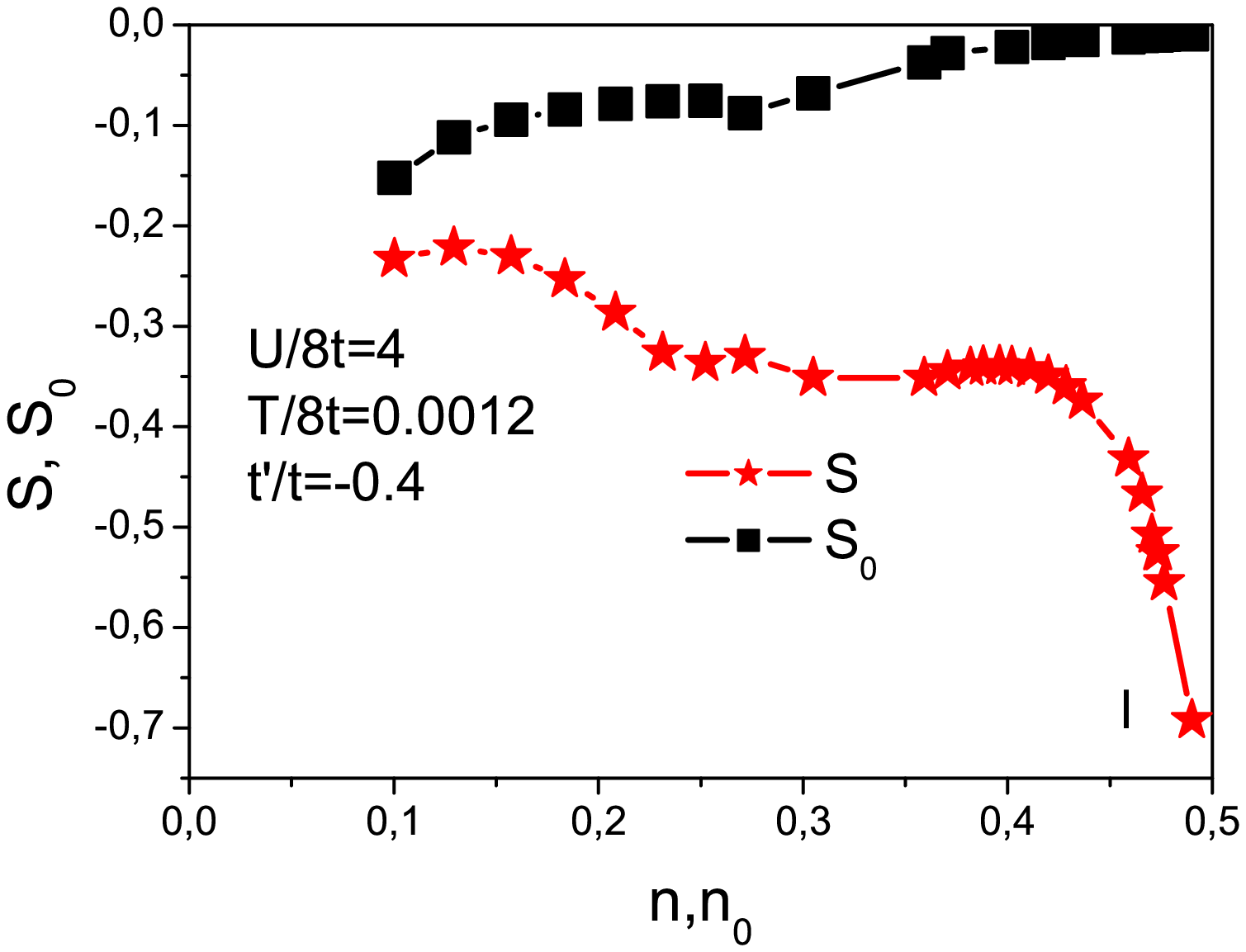}
\caption{Relation between the effective band -- filling of the band without
correlations $n_0$ and total band -- filling of correlated band
$n$ (a,d,g,j). Comparison of estimates for Hall coefficient $R_{H0}$ (b,e,h,k)
and thermopower $S_0$ (c,f,i,l) with DMFT results ($R_H$, $S$).}
\label{fig8}
\end{figure}

In Fig. \ref{fig8} we show relations between $n_{0}$ and $n$, as well as
simple estimates for Hall coefficient and thermopower, compared with exact
DMFT results. In high -- temperature regime ($T/8t=0.0586$), when quasiparticle
peak in the density of states vanishes, we can see that Eq. (\ref{n_n0}) is
more or less confirmed. Estimates for thermopower $S_{0}$ and especially for
Hall coefficient obtained in this approach are also sufficiently close to exact
results.

In low -- temperature regime the presence of quasiparticle peak in the density
of states of DMFT makes Eq. (\ref{n_n0}) invalid and $n_{0}$ in fact just
coincides with  $n$. The estimate for Hall coefficient $R_{H0}$ is still quite
close to exact DMFT results (except anomalies in $R_{H}$, related to
quasiparticle peak for $t'=0$ \cite{KKKS1,KKKS2} (Fig. \ref{fig8}h). At the
same time the estimate of thermopower at low temperatures is unsatisfactory.
It is possibly related to more important role of ``smearing'' of spectral
density around the Fermi surface in calculations of thermopower and anomalies
related to filling of quaiparticle peak.

It is necessary to note once more, that anomalies observed in low -- temperature
regime are determined by the presence of quasiparticle peak in DMFT
approximation, while it vanishes with increase of temperature or due to disorder,
as well as due to account of nonlocal correlations outside DMFT. Thus, it is
quite possible that qualitatively the results obtained in high -- temperature
regime are more or less general.

Using ARPES data for doped Mott insulator we can determine
$n_{0}$ (in fact it is just the area of the Brillouin zone below the Fermi
surface) and then, using Eq. (\ref{n_n0}), define the total filling $n$ and
hole doping level as $p=1-2n$. Standard DFT calculations of electron spectrum,
with the account of filling of uncorrelated band $n_{0}$, allow using
Eqs.  (\ref{S5}) and (\ref{R_H3}) to estimate both $R_{H}$ and $S$.
In fact these results are in total agreement with the picture of ``hidden''
Fermi -- liquid introduced in Ref. \cite{Kot}.

\section{Conclusion}

We have studied the behavior of thermoelectric power in metallic phase
originating by doping the Mott insulator. We mainly concentrated on the case of
hole doping, characteristic for the major part of cuprates. We have considered
a number of two -- dimensional tight -- binding models of electron spectrum,
fitting the electronic structure of cuprates. In all of these models in doped
Mott insulator we observe anomalous temperature dependence of thermopower $S$
which is significantly different from linear dependence typical to usual metals.
In low -- temperature limit $S$ is mainly determined by filling of quasiparticle
peak, leading to negative values of thermopower at any hole doping and anomalous
dependence of $S$ on filling, when the negative (electron -- like)
thermopower increases its absolute value with addition of electrons.
For thermopower these low -- temperature anomalies, connected with the filling
of quasioarticle peak, are much stronger than similar anomalies for Hall
coefficient. In high -- temperature limit, when the quasiparticle peak is
essentially damped, $S$ is mainly determined by the filling of lower (for hole
doping) or upper (for electron doping) Hubbard band. In this limit the
qualitative estimate shows that the sign change of both $S$ and Hall effect in
the simplest (symmetric, $t'=0$) case takes place at band -- filling $n$=1/3
per single spin projection, corresponding to hole doping $p=1-2n$=1/3.


DMFT calculations show that $S$ changes its sign at $n$=0.42, while Hall effect
sign change takes place at $n$=0.36. Thus, even in the case of electron --
hole symmetry there is narrow region of dopings, where $R_H$ is already positive,
while $S$ remains negative. The breaking of electron -- hole symmetry leads to
the increase of filling $n$, corresponding to sign change of $S$, and to the
decrease of $n$, corresponding to sign change of $R_H$, leading to rather wide
region of fillings, where thermopower and Hall effect have different signs.
We have also proposed a scheme, allowing to obtain the number of charge
carriers in doped Mott insulator form ARPES data using DFT calculations of
electronic spectrum and to perform a semiquantitative estimate of both Hall
coefficient and thermopower.

In principle, on the qualitative level, the possibility of different signs of
thermopower and Hall effect is known for rather long time and was observed in a
number of experiments in disordered systems \cite{Allg,MD}. Systematic studies
of thermopower in cuprates for different doping levels was performed in
Refs. \cite{OCT,HH}. It was demonstrated that thermopower of a number of
cuprates changes its sign close to $p=1-2n\approx$0.1--0.2, in the vicinity of
optimal doping level $p\approx$ 0.16, corresponding to the maximum of
superconducting temperature. In Ref. \cite{ASDP} this behavior of thermopower
was interpreted as connected to the presence of a nearby quantum critical
point, related with separation of the upper Hubbard band. Unfortunately, no
comparison was made in these works with the available data on Hall effect.
In Refs. \cite{Boeb,Tal1,Tal2,PrTal} the measurements of Hall effect were done
for several cuprates at low temperatures in extremely strong magnetic fields
(in the normal phase), which also demonstrated the anomalous behavior
(growth of the Hall number of carriers) at doping levels $p\approx$0.2--0.25.
I these works this anomalous behavior of Hall effect was also related to the
nearby quantum critical point, corresponding to the closure of the pseudogap.
In Refs. \cite{KKKS1,KKKS2} this behavior of Hall effect was interpreted as
due to the approach to the point, where Hall effect changes its sign (which
actually was not observed in the samples studied in
Refs.  \cite{Boeb,Tal1,Tal2,PrTal}), with no relation to any quantum critical
point. The DMFT results obtained above are in qualitative agreement with
experimental data. In particular, the hole doping level of Mott insulator,
corresponding to the change of sign of the thermopower is always lower, than
the doping level where the Hall coefficient changes its sign. This stresses the
importance of systematic comparative studies of Hall effect and thermopower in
cuprates.

Authors are grateful to P. Phillips, who called our attention to the
importance of the studies of concentration dependence of thermopower within
DMFT, and to D.I. Khomskii who actively participated in the initial stages of
our work on the Hall effect.


\end{document}